\documentclass[aps,prd,twocolumn,groupedaddress,floatfix,nofootinbib,showpacs]{revtex4}
\newcommand{\twocol}[2]{#1}
\usepackage{graphicx,amsmath,amssymb,amsfonts,dcolumn,setspace,longtable}
\twocol{\setlength{\LTcapwidth}{.85\textwidth}}{\setlength{\LTcapwidth}{\textwidth}}

\begin{document}

\title{Exploring the meson spectrum with twisted mass lattice QCD}

\author{Robert G.\ Petry$^a$, Derek Harnett$^b$, Randy Lewis$^{a,c}$,
        R. M. Woloshyn$^d$}
\affiliation{\vspace{2mm}
             $^a$Department of Physics, University of Regina, Regina, SK,
                 S4S 0A2, Canada \\
             $^b$Department of Physics, University College of the Fraser Valley,
                 Abbotsford, BC, V2S 7M8, Canada \\
             $^c$Department of Physics and Astronomy, York University, Toronto,
                 ON, M3J 1P3, Canada \\
             $^d$TRIUMF, 4004 Wesbrook Mall, Vancouver, BC, V6T 2A3, Canada}

\date{March 28, 2008}

\begin{abstract}
Numerical simulations with access to all possible meson quantum numbers,
$J^{PC}$, are presented using two-flavor (up and down) quenched twisted mass
lattice QCD with three different lattice spacings and four different quark
masses.  The connection between the quantum numbers ($P$ and $C$) and the
symmetries of the twisted mass
action is discussed, as is the connection between $J$ and the lattice rotation
group, for the 400 operators used in this study.  Curve fitting of this large
data set is accomplished by using an evolutionary fitting algorithm.
Results are reported for conventional and exotic quantum numbers.

%%% Local Variables:
%%% mode: latex
%%% TeX-master: "tmmeson"
%%% End:

\end{abstract}

 % Interpolation; curve fitting

%\maketitle must follow title, authors, abstract, \pacs, and \keywords
\maketitle

\newcommand{\python}{\textsc{Python}}
\newcommand{\irrepdim}{d}
\newcommand{\order}[1]{g_{\scriptscriptstyle{{#1}}}}
\newcommand{\elem}{R}
\newcommand{\Ogroup}{O}
\newcommand{\class}{\ensuremath{\xi}}
\newcommand{\character}{\ensuremath{\chi}}
\newcommand{\oldirrep}{\ensuremath{\Gamma}}
\newcommand{\rep}[4]{\ensuremath{\Gamma}^{({#1})}_{{#2}{#3}}({#4})}
\newcommand{\repmatrix}[2]{\ensuremath{\Gamma}^{({#1})}({#2})}
\newcommand{\reponly}[1]{\ensuremath{\Gamma}^{({#1})}}
\newcommand{\irrepcopy}{{\ensuremath{\alpha}}}
\newcommand{\irrepone}{{\ensuremath{\Lambda_1}}}
\newcommand{\irreptwo}{{\ensuremath{\Lambda_2}}}
\newcommand{\irrep}{\ensuremath{\Lambda}}
\newcommand{\irreponei}{{\ensuremath{\lambda_1}}}
\newcommand{\irreptwoi}{{\ensuremath{\lambda_2}}}
\newcommand{\irrepi}{\ensuremath{\lambda}}
\newcommand{\irreponej}{{\ensuremath{\mu_1}}}
\newcommand{\irreptwoj}{{\ensuremath{\mu_2}}}
\newcommand{\irrepj}{\ensuremath{\mu}}
\newcommand{\link}{\ensuremath{U}}
\newcommand{\linklower}{\ensuremath{u}}
\newcommand{\irreplink}{\ensuremath{\Lambda_\linklower}}
\newcommand{\irreplinki}{\ensuremath{\lambda_\linklower}}
\newcommand{\dirac}{\ensuremath{F}}
\newcommand{\diraclower}{\ensuremath{f}}
\newcommand{\irrepdirac}{\ensuremath{\Lambda_\diraclower}}
\newcommand{\irrepdiraci}{\ensuremath{\lambda_\diraclower}}
\newcommand{\spinor}{\ensuremath{\psi}}
\newcommand{\spinorbar}{\ensuremath{\overline{\spinor}}}
\newcommand{\mesonop}{\ensuremath{M}}
\newcommand{\myvector}[1]{\ensuremath{\mathbf{{#1}}}}
\newcommand{\vecx}{\myvector{x}}
\newcommand{\unit}[1]{\ensuremath{\hat{{#1}}}}
% The following define macros for the matrix representations
\newcommand{\cyaone}{\ensuremath{\left[\begin{array}{c}
1
\end{array}\right]
}}
\newcommand{\czaone}{\ensuremath{\left[\begin{array}{c}
1
\end{array}\right]
}}
\newcommand{\cyatwo}{\ensuremath{\left[\begin{array}{c}
-1
\end{array}\right]
}}
\newcommand{\czatwo}{\ensuremath{\left[\begin{array}{c}
-1
\end{array}\right]
}}
\newcommand{\cye}{\ensuremath{\frac{1}{2}\left[\begin{array}{cc}
1 & \sqrt{3} \\
\sqrt{3} & -1 
\end{array}\right]
}}
\newcommand{\cze}{\ensuremath{\left[\begin{array}{cc}
-1 & 0 \\
0 & 1 
\end{array}\right]
}}
\newcommand{\cytone}{\ensuremath{\left[\begin{array}{ccc}
0 & 0 & 1 \\
0 & 1 & 0 \\
-1 & 0 & 0 
\end{array}\right]
}}
\newcommand{\cztone}{\ensuremath{\left[\begin{array}{ccc}
0 & -1 & 0 \\
1 & 0 & 0 \\
0 & 0 & 1 
\end{array}\right]
}}
\newcommand{\cyttwo}{\ensuremath{\left[\begin{array}{ccc}
0 & 1 & 0 \\
-1 & 0 & 0 \\
0 & 0 & -1 
\end{array}\right]
}}
\newcommand{\czttwo}{\ensuremath{\left[\begin{array}{ccc}
-1 & 0 & 0 \\
0 & 0 & 1 \\
0 & -1 & 0 
\end{array}\right]
}}

% Main text

\section{\label{sec:intro}Introduction}

Although many mesons have been observed in nature, the spectrum of known
mesons does not include all the states expected in Quantum Chromodynamics.
Experimental searches and theoretical studies are
continuing, but gaps in current knowledge leave room for new
approaches.  Lattice QCD is an established method for extracting numerical
predictions directly from the underlying quantum field theory, but lattice
explorations of the full spectrum of mesons still suffer from limitations.
For reviews of both theory and experiment for light-quark mesons, see
Refs.~\cite{Godfrey:1998pd,Close:2001zp,Klempt:2007cp}.

A helpful framework for beginning the discussion of light-quark mesons is
provided
by the constituent quark model, where mesons are considered to be composed
of a system of two bound quarks whose spins can couple to a singlet
($S=0$) or triplet ($S=1$) total spin which in turn can couple with
relative angular momentum $L$ between the quarks to produce a total
observed angular momentum $J$.  Using spectroscopic notation for the
states $^{2S+1}L_J$, one obtains the familiar list of accessible $J^{PC}$,
as shown in Table~\ref{tab:quarkmodel}.

Since in QCD gluon fields are also present, it is possible for
gluonic excitations in mesons to contribute non-trivially to
the observed quantum numbers of a meson.  Such \emph{hybrid
 mesons} with the same $J^{PC}$ as conventional mesons are difficult
to distinguish but, as shown in Table~\ref{tab:quarkmodel}, there is a subclass
of hybrid mesons, called \emph{exotic mesons}, with quantum numbers
unattainable in the quark model, eg.
$0^{--}$, $0^{+-}$, $1^{-+}$, $2^{+-}$, and $3^{-+}$.
These exotic mesons, for which there is yet no definitive experimental evidence,
would offer a clear signature for excited gluon dynamics.

\newcolumntype{g}{D{.}{.}{3}}
\begin{table}
\caption{\label{tab:quarkmodel} States accessible in the constituent quark
 model for different total spin $S$ and orbital angular momentum $L$ labeled by
 spectroscopic notation $^{2S+1}L_J$ with corresponding $J^{PC}$.}
\begin{ruledtabular}
\begin{tabular}{ggggl}
  & \multicolumn{2}{l}{~~~~$S=0$ (singlet)} & \multicolumn{2}{l}{~~~~~~~$S=1$ (triplet)}\\
L & ^{2S+1}L_J & J^{PC} & ^{2S+1}L_J & ~~~~~~$J^{PC}$ \\ \hline
S & ~~^1S_0 & 0^{-+} & ~~^3S_1 & $1^{--}$ \\ 
P & ~~^1P_1 & 1^{+-} & ~~^3P_J & $0^{++},~1^{++},~2^{++}$ \\
D & ~~^1D_2 & 2^{-+} & ~~^3D_J & $1^{--},~2^{--},~3^{--}$ \\
F & ~~^1F_3 & 3^{+-} & ~~^3F_J & $2^{++},~3^{++},~4^{++}$
\end{tabular}
\end{ruledtabular}
\end{table}

Twisted mass lattice QCD (tmLQCD)~\cite{Frezzotti:2003ni,Frezzotti:2000nk} 
is used for the lattice simulations in this work.  
It is not clear {\em a priori} whether tmLQCD is a very favorable 
action for lattice simulations of the full meson spectrum.
The fact that tmLQCD does not respect parity $P$ may be seen as an
obstacle, but we will show that it is not insurmountable.
Meanwhile, tmLQCD has the advantages of offering a cost-effective approach to
the chiral limit as well as a simple removal of the leading, i.e.\ $O(a)$,
lattice spacing errors.
In this work, it is shown that tmLQCD's reduced symmetries can be understood
and accounted for in practical simulations, thereby allowing the reader to make
an informed decision as to whether this option is preferable for
future studies of the meson spectrum.
The issues specific to tmLQCD that affect our operators are discussed in
Section~\ref{sec:twistedmass}.

Much of the discussion of meson operators is not specific to tmLQCD.
See Section~\ref{sec:latticeops} for this tmLQCD-independent presentation.
In particular, we choose meson correlators that can be built from quark and
anti-quark propagators originating at a single lattice site augmented by 
gauge field links defined on extended spatial paths.

All possible quantum numbers are obtained by using a variety of options for the
paths of gauge fields connecting quark to anti-quark.
(Gauge fields are subsequently smeared at both source and sink, but quarks are
only smeared at the sink in this work.)
With just one quark propagator inversion, these ``excited glue'' operators are
minimally expensive, and the results of our simulations allow us to tabulate
the relative strengths of the overlaps that various mesons have with
various operators.  It is interesting to see the extent to which these
 ``excited glue'' operators couple to exotic mesons, and also to conventional
quantum numbers.

Meson masses are extracted from lattice QCD simulations by fitting a
linear combination of exponentials to correlators as a function of
Euclidean time. It is a delicate business.
For example, the number of exponentials that should be used for a certain fit
depends on the particular channel being studied and also on the quality and
quantity of data.  Our study of the meson spectrum will include hundreds
of correlators, some of which should be fit simultaneously since each meson will
appear in multiple correlators.  Because sink smearing is more easily varied
than source smearing due to the expense of recomputing quark propagators, we
want a fitting method that does not require computation of a complete
correlator matrix, i.e.\ we want the freedom to consider more sink options
than source options.

To address all of the delicate issues of fitting, the evolutionary algorithm 
introduced in Refs.~\cite{von_Hippel:2007ar,vonHippel:2007dz} is used.
The workings of the algorithm are
well-understood in terms of the basic principles underlying biological
evolution, but it is a black box algorithm in the sense that human
intervention, and therefore human bias, is avoided.
For example, the algorithm will identify the number of exponentials that
minimize the $\chi^2/n_{dof}$ for a given fit.
This black box method is also general enough to handle multi-correlator fits
with no need of a complete correlator matrix.
The data-fitting technique proposed in
Refs.~\cite{von_Hippel:2007ar,vonHippel:2007dz} is independent of tmLQCD.
This is its first application to such a large set of lattice QCD data, and
the implementation is described in Section~\ref{sec:curvefit}.

Section~\ref{sec:simdetails} explains the parameter choices used in our
numerical simulations.
The present study is exploratory, and we have therefore chosen to perform
quenched simulations.\footnote{For the status of dynamical tmLQCD, see
Ref.~\cite{Boucaud:2008xu} and references therein.}
Lessons learned in this study about extended meson
operators, fitting algorithms, tmLQCD and the meson spectrum are applicable to
future studies beyond the quenched approximation.
Section~\ref{sec:results} presents and discusses the results.

Section~\ref{sec:conclusion} draws some conclusions about the meson spectrum.
It also highlights properties of our chosen operators, comments on
the appropriateness of the twisted mass action to this physics, and underscores
the valuable qualities of the evolutionary fitting technique.

%%% Local Variables: 
%%% mode: latex
%%% TeX-master: t
%%% End: 

\section{\label{sec:latticeops}Lattice-symmetrized meson operators}

Lighter quark mass calculations require improved statistics for hadron
mass resolution.  This requirement and the need to disentangle
physical states with the same quantum numbers may be accomplished 
in part through the creation of more operators that represent the
channel in question.  The creation of more elaborate operators also
allows for the study of hybrid and exotic mesons.

Operators with displaced quarks have been constructed using group
theoretical techniques but their usage requires the calculation of
quark propagators from multiple lattice sites.  (See, for example,
Refs.~\cite{Lacock:1996vy,Basak:2005aq,Basak:2005ir}.)  Consideration of the
operators available through gluonic extension alone therefore is
numerically expedient.  Moreover, it allows investigation of the
coupling of operators with ``excited glue'' to conventional and exotic
states.  This section provides a complete discussion of the
operators (first introduced in
Ref.~\cite{Harnett:2006fp}) used in our simulation.

\subsection{Lattice symmetry group}

While parity ($P$) and charge conjugation ($C$) may be conserved by
lattice actions, the continuous rotational symmetry of nature is
broken and one requires operators adapted to the symmetry group of the
lattice.  For mesons this is the octahedral group $\Ogroup$ with $24$
elements. The group $O$ has five conjugacy classes conventionally
labeled $\{E,3C_4^2,8C_3,6C_4,6C_2\}$ and therefore admits five
(unitary) irreducible representations (irreps): two one-dimensional irreps $A_1$
and $A_2$, one two-dimensional irrep $E$, and two three-dimensional irreps
$T_1$ and $T_2$.
The direct product of the parity, charge conjugation, and octahedral
groups is denoted $\Ogroup^{PC}$.

For an operator adapted to the representation $\irrep^{PC}$, where
\mbox{$\irrep\in\{A_1,A_2,E,T_1,T_2\}$} is an irrep of $\Ogroup$ and
\mbox{$P,C\in\{+,-\}$}, one identifies the possible physical states
$J^{PC}$ to which it corresponds using
Table~\ref{tab:irrepcorrespondence} which shows the number of copies
%$n^J_\irrep$ of irrep $\irrep$ to which the continuum $SO(3)$ irrep
%$J$ subduces~\cite{Johnson:1982yq}.  For example, an operator
$n^J_\irrep$ of irrep $\irrep$ contained in the reduction of the subduced continuum rotation group ($SO(3)$) irrep
$J$~\cite{Johnson:1982yq}.  For example, an operator
transforming as the irrep $E^{+-}$ could have spin content
\mbox{($J^{PC}=2^{+-}$, $4^{+-}$, $5^{+-}$, $\ldots$)}. In theory one
needs many operators of each $\irrep^{PC}$ to resolve the range of
physical spins $J^{PC}$ in the tower of states to which $\irrep^{PC}$
corresponds.

\begin{table}
\caption{\label{tab:irrepcorrespondence}Number of copies
    $n^J_\irrep$ of the irrep $\irrep$ of $\Ogroup$ in the reduction
    of the subduced representation of the continuum irrep $J$ of
    $SO(3)$.}
%  Subduction of continuum irreps $J$ of $SO(3)$ to those of the lattice symmetry group $\Ogroup$.}
$
\begin{array}{|r|rrrrr||r|rrrrr|} \hline
J & A_1 & A_2 & E & T_1 & T_2 &
J & A_1 & A_2 & E & T_1 & T_2 \\ \hline
0 & 1 & 0 & 0 & 0 & 0 &
3 & 0 & 1 & 0 & 1 & 1 \\
1 & 0 & 0 & 0 & 1 & 0 & 
4 & 1 & 0 & 1 & 1 & 1 \\
2 & 0 & 0 & 1 & 0 & 1 & 
5 & 0 & 0 & 1 & 2 & 1 \\ \hline
\end{array}
$
\end{table}

Table~\ref{tab:irrepcorrespondence} is arrived at through a
consideration of the character table of $\Ogroup$ shown in
%Table~\ref{tab:ochartable}.  The trace of the $SO(3)$ matrix for spin
Table~\ref{tab:ochartable}.  The trace of the rotation group matrix for spin
$J$ and rotation angle $\theta(\class)$ corresponding to class
$\class$ is given by\footnote{This formula is to be interpreted as a
  limit in the event the denominator vanishes.}
\begin{equation}
\label{eq:continuumchar}
\character^{(J)}(\class)\ = \sin[(J+1/2)\theta(\class)]/\sin[\theta(\class)/2]\;.
\end{equation}
Those matrices in the continuous irrep $J$ which correspond to the
subgroup $O$ of $SO(3)$ form a representation of $O$ which is, in
general, now reducible.  Equation~(\ref{eq:continuumchar}) combined
with Table~\ref{tab:ochartable}, allows one to determine the
subduction of continuum $J$ to discrete $\irrep$ of $\Ogroup$ via the
decomposition formula
\begin{equation}
\label{eq:2}
n^J_\irrep=\frac{1}{\order{\Ogroup}}
\sum_\class p_\class \character^{(\irrep)}(\class)^*\character^{(J)}(\class)
\end{equation}
for the number $n^J_\irrep$ of copies of $\irrep$ in the subduction of
$J$.  Here $p_\class$ is the number of elements in class $\class$ and
$\order{\Ogroup}$ is the order of the octahedral group.

\begin{table}
\caption{\label{tab:ochartable}Character table of $\Ogroup$ showing the character $\character$ for the given class $\class$ in the irrep $\irrep$.  The final line displays the angle of rotation $\theta$ common to the elements in each class required in Equation~(\ref{eq:continuumchar}).}
$
\begin{array}{|c|rrrrr|} \hline
\irrep\backslash\class & E & 3C_4^2 & 8C_3 & 6 C_4 & 6 C_2 \\ \hline
A_1 & 1 & 1 & 1 & 1 & 1 \\
A_2 & 1 & 1 & 1 & -1 & -1 \\
E & 2 & 2 & -1 & 0 & 0 \\
T_1 & 3 & -1 & 0 & 1 & -1 \\
T_2 & 3 & -1 & 0 & -1 & 1 \\ \hline
\theta(\class) & 2\pi & \pi & 2\pi/3 & \pi/2 & \pi \\ \hline
\end{array}
$
\end{table}

\subsection{Operator building blocks}
\begin{figure}
\includegraphics[width=.2\textwidth,clip=]{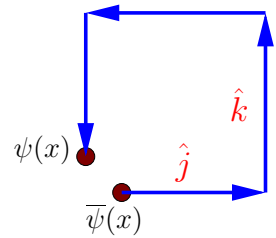}
\caption{Diagrammatic representation of the building block operator.}
\label{fig:buildingblock}
\end{figure}

One may construct zero-momentum operators transforming as irreps of
$\Ogroup^{PC}$ from the space of operators spanned by
%R The following would work in thesis, but not in two-column
%\begin{equation}
%\label{eq:1}
%M_{j,k,a,b}(t)=\sum_\mathbf{x}\spinorbar_a(x)
% U_j(x)U_k(x+\unit{j})U_{-j}(x+\unit{j}+\unit{k})U_{-k}(x+\unit{k}) \spinor_b(x)\ ,
%\end{equation}
\begin{equation}
\label{eq:1}
M_{j,k,a,b}(t)=\sum_\mathbf{x}\spinorbar_a(x)
 U_{j,k}(x)\spinor_b(x)\ ,
\end{equation}
where the gauge link part of our operators is defined via
\begin{equation}
U_{j,k}(x)\equiv
U_j(x)U_k(x+\unit{j})U_{-j}(x+\unit{j}+\unit{k})U_{-k}(x+\unit{k})\
.
\end{equation}
Here $j,k = \pm 1,\pm 2,\pm 3; j \ne k$ and $a$ and $b$ are spinor
indices for a total of $24\times16=384$ operators.  The $\hat{j}$
denotes a four-vector of unit length along the spatial axis~$j$.  See
Figure~\ref{fig:buildingblock} for a diagrammatic representation of
the building block operator.
Superpositions of such operators for meson spectrum analysis have been
suggested and used previously in special
cases~\cite{Mandula:1983wc,Jorysz:1987qj,Lacock:1996vy,Bernard:1997ib}.
Evaluation of correlators constructed from operators of this form
requires only calculation of propagators from a single source.
Construction of operators transforming as irrep $\irrep^{PC}$ is
facilitated by observing that one can effectively consider the
transformation of the spinor and link paths independently and then
combine them via octahedral Clebsch-Gordan coefficients.  Link
smearing can be performed on operators at both the source and sink of
the correlator, and quark smearing at one end to suppress high energy
states with no change to the symmetry of the symmetrized operators.

\subsection{Spinor contribution to group structure}
\label{sec:spin-contr-group}
The contribution to the group structure due to spinor indices is
determined by the $16$ bilinears $\spinorbar\dirac
\spinor$, where $\dirac$ represents one of sixteen $4\times4$ matrices,
\mbox{$\{I,\gamma_5,\gamma_4,\gamma_4\gamma_5,
  \gamma_i,\gamma_i\gamma_5,\sigma_{4i},\epsilon_{ijk}\sigma_{jk}\}$}.
The first four bilinears are scalars while the last four three-index
objects are vectors under the rotation group in the Euclidean
continuum.

Parity ($P$) and charge conjugation ($C$) of the bilinears are
identified via
\begin{equation}\label{PandC}
\begin{array}{ccc}
\mathcal{C}\spinor\mathcal{C}^\dagger = \left(\spinorbar C^\dagger\right)^T, &
\hspace{1em} &
\mathcal{P}\spinor\mathcal{P}^\dagger = \gamma_4 \spinor, \\
\mathcal{C}\spinorbar\mathcal{C}^\dagger = -\left(C \spinor\right)^T, &
&
\mathcal{P}\spinorbar\mathcal{P}^\dagger = \spinorbar \gamma_4,
\end{array}
\end{equation}
where $C$ is the matrix implementing charge conjugation and we have
suppressed the action on coordinates, which under parity sees
$x=(x_1,x_2,x_3,x_4)\to x_P$, where
\begin{equation}\label{xP}
x_P = (-x_1,-x_2,-x_3,x_4).
\end{equation}

To classify the bilinears into irreps of $\Ogroup$ one uses its
character table found in Table \ref{tab:ochartable} to project out the
octahedral irreps from the representations generated by the bilinears.
Since the bilinears are scalars and vectors in the continuum it
follows that their respective spans are also invariant under
$\Ogroup$.  By inspection it is straightforward to show that the
character table for the scalar and vector bilinears is identical with
that of $A_1$ and $T_1$ respectively.  Since the multiplicity
$n^J_\irrep$ of irrep $\irrep$ in \emph{any representation} $J$ is
given by Equation~(\ref{eq:2}), it follows trivially that the scalar
and vector bilinears form the basis of $A_1$ and $T_1$ irreps
respectively.

Furthermore it may be verified that the $i^{th}$ vector bilinear
component transforms as the $i^{th}$ row for our choice of matrix
representation $\repmatrix{T_1}{R}$ given in
Table~\ref{tab:irrepmatrices}.  Since the action on the spinor under
rotation $R$ is $\spinorbar\rightarrow\spinorbar S(R)$ and
$\spinor\rightarrow S^\dagger(R)\spinor$, this amounts to verifying
that
\begin{equation}
\label{eq:diracrotation}
S(R)\dirac_i S^\dagger(R)=\sum_j \rep{T_1}{j}{i}{R}\dirac_j\;,
\end{equation}
for the elements $R$ of $\Ogroup$ and the four different vector
bilinears $F$.  Here $S(R)$ is the matrix implementing the rotation on
the spinors.  It is sufficient to verify
Equation~(\ref{eq:diracrotation}) for the two generators, namely the
$\pi/2$ rotations about the $y$ and $z$ axes, $C_{4y}$ and $C_{4z}$, for
which these matrices are given by
\begin{eqnarray}
S(C_{4y}) & = &\frac{1}{\sqrt{2}}(1+\gamma_1\gamma_3)\;, \\
S(C_{4z}) & = &\frac{1}{\sqrt{2}}(1+\gamma_2\gamma_1)\;.
\end{eqnarray}

\begin{table}
\begin{ruledtabular}
\caption{\label{tab:irrepmatrices}Matrices $\Gamma$ for the generators $C_{4y}$ and $C_{4z}$ of
  the octahedral group for each irrep $\Lambda$.}
\begin{tabular}{lcc}
$\irrep$ & $\repmatrix{\irrep}{C_{4y}}$ & $\repmatrix{\irrep}{C_{4z}}$ \\ \hline 
$A_1$ \rule[-2.3ex]{0ex}{5.4ex} & $\cyaone$  & $\czaone$ \\
$A_2$ \rule[-2.3ex]{0ex}{4.6ex}& $\cyatwo$ & $\czatwo$ \\
$E$   \rule[-4ex]{0ex}{8ex}& $\cye$    & $\cze$    \\
$T_1$ \rule[-5ex]{0ex}{10ex}& $\cytone$ & $\cztone$ \\
$T_2$ \rule[-5ex]{0ex}{10ex}& $\cyttwo$ & $\czttwo$
\end{tabular}
\end{ruledtabular}
\end{table}

Hence the reduction of the spinor structure of our operators is given
in Table \ref{tab:spinorreduction}, where now each bilinear may be
classified uniquely by its irrep $\irrep^{PC}$, row~$\irrepi$, and
irrep multiplicity index~$\alpha$ as $\dirac^{\irrep^{\irrepcopy
    PC}}_\irrepi$.

\begin{table}
\caption{\label{tab:spinorreduction}Octahedral symmetries of the spinor bilinears.  The operators have been separated into quadrants according to their dependency on twist angle as discussed in Ref.~\protect\cite{Harnett:2006fp}.}
$
\begin{array}{|c|c|c|c||c|c|c|c|} \hline
\dirac & \rule{0ex}{2.5ex}\irrep^{PC} & \irrepi & \irrepcopy &
\dirac & \rule{0ex}{2.5ex}\irrep^{PC} & \irrepi & \irrepcopy \\ \hline\hline
{I} & \rule{0ex}{2.5ex}{A_1^{++}} & {1} & {1} &
{\gamma_i} & \rule{0ex}{2.5ex}{T_1^{--}} & {i} & {1} \\
{\gamma_5} & \rule{0ex}{2.5ex}{A_1^{-+}} & {1} & {1} &
{\gamma_i\gamma_5} & \rule{0ex}{2.5ex}{T_1^{++}} & {i} & {1} \\ \hline
{\gamma_4} & \rule{0ex}{2.5ex}{A_1^{+-}} & {1} & {1} &
{\sigma_{4i}} & \rule{0ex}{2.5ex}{T_1^{--}} & {i} & {2} \\
{\gamma_4\gamma_5} & \rule{0ex}{2.5ex}{A_1^{-+}} & {1} & {2} &
{\epsilon_{ijk}\sigma_{jk}} & \rule{0ex}{2.5ex}{T_1^{+-}} & {i} & {1} \\ \hline 
\end{array}
$
\end{table} 

\subsection{Link contribution to group structure}
\label{sec:link-contr-group}
\begin{figure}
\includegraphics[width=\columnwidth,clip=]{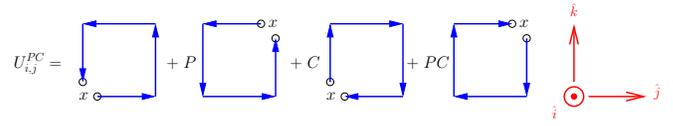}
\caption{$PC$-symmetrized basis elements for gauge field links.  We
  have suppressed the overall normalization factor of $1/2$.}
\label{fig:upc}
\end{figure}

As with the spinors, one can simplify the discussion of the rotational
properties of the gauge links by considering first the parity and
charge conjugation and only then the rotations in $\Ogroup$. Parity of
a link about the point $x$ is implemented via inversion about all
three spatial axes as usual.  Charge conjugation sees
$U\rightarrow U^*$, however to compensate for an overall transpose
arising from the transformation of our quark bilinear, we formally
transform $U\rightarrow U^\dagger$ to achieve the correct overall
charge conjugation properties of our operators.  The parity and charge
conjugation contributions due to this link structure can then be taken
into account by defining the $PC$-adapted superpositions,
\begin{equation}
\label{eq:pcadaptedu}
U^{PC}_{i,j} = \frac{1}{2}\left(U_{j,k} + P U_{-j,-k} + C U_{k,j} + PC U_{-k,-j}\right)\;,
\end{equation}
where $k$ is defined via $\hat{k}=\hat{i}\times\hat{j}$.
Diagrammatically, $U^{PC}_{i,j}$ is shown in Figure~\ref{fig:upc}.
For fixed $P$ and $C$ the space spanned by $U^{PC}_{i,j}$ is invariant
under~$\Ogroup$ and will generate a representation $U^{PC}$ of the
group.  A basis for the six-dimensional space may be found by restricting $(i,j)$
to $\{(1,2),(1,3),(2,3),(2,1),(3,1),(3,2)\}$ and this will in turn
induce the particular matrix representation
$\rep{U^{PC}}{(i,j)}{(m,n)}{\elem}$.

In order to determine how each of these four six-dimensional representations
reduces, it is convenient to calculate their respective character
tables.  A straightforward consideration of the effect of a single
element from each class of $\Ogroup$ on the basis vectors allows one
to find the character table for $U^{PC}$ shown in
Table~\ref{tab:upcchartable}.
\begin{table}
\caption{Character table showing $\character^{(U^{PC})}(\class)$ for representation $U^{PC}$ induced by $U^{PC}_{i,j}$.  Here positive and negative are to be interpreted as $+1$ and $-1$ in the entries.}
\label{tab:upcchartable}
$
\begin{array}{|ccccc|} \hline
E & 3C_4^2 & 8C_3 & 6 C_4 & 6 C_2 \\ \hline
6 & 2P & 0 & 0 & (P+1)C \\ \hline
\end{array}
$
\end{table}
Use of Equation (\ref{eq:2}) with $J$ as $U^{PC}$ reduces the latter to
irreps $\irrep^{PC}$ of $\Ogroup^{PC}$ as shown in
Table~\ref{tab:upcreduction}.  So, for example, $U^{++}=A_1^{++}\oplus
E^{++}\oplus T_2^{++}$.

\begin{table}
\caption{Reduction of $U^{PC}$ to irreps $\irrep^{PC}$ of
  $\Ogroup^{PC}$.}
\label{tab:upcreduction}
$
\begin{array}{|cc|ccccc|} \hline
P & C & \rule{0ex}{2.5ex}A_1^{PC} & A_2^{PC} & E^{PC} & T_1^{PC} & T_2^{PC} \\ \hline
+ & + & 1 & 0 & 1 & 0 & 1 \\
+ & - & 0 & 1 & 1 & 1 & 0 \\
- & + & 0 & 0 & 0 & 1 & 1 \\
- & - & 0 & 0 & 0 & 1 & 1 \\ \hline
\end{array}
$
\end{table}

\begin{figure}
\includegraphics[width=.8\columnwidth,keepaspectratio=,clip=]{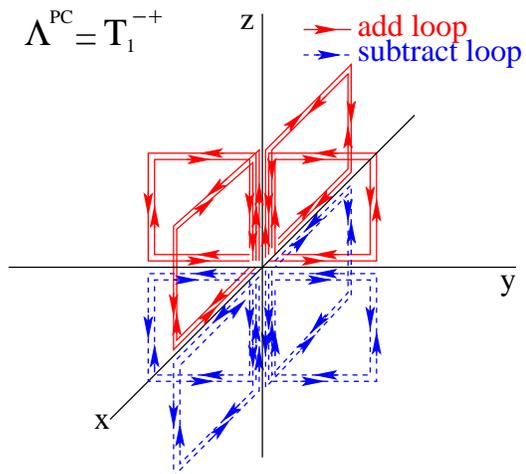}
\caption{\label{fig:Ut1mp} Extended gauge structure
  $\link^{\irrep^{PC}}_{\irrepi}=\link^{T_1^{-+}}_3$ is shown.  Parity
  and charge conjugation symmetries are readily verified.  The $T_1$
  irrep has three rows, this being the $\irrepi=3$ component which is
  symmetric about the $z$-axis.  Illustrations of other
  lattice-symmetrized gauge structures may be found in
  Ref.~\cite{vonHippel:2007dz}.  }
\end{figure}

To arrive at the explicit form of the lattice-symmetrized gauge links,
note that, by Table~\ref{tab:upcreduction}, each of our spaces
reduces to no more than a single copy of each irrep.  For such
\emph{simply reducible} representations the reduction is
straightforward\footnote{See Ref.~\cite[page 74]{Cornwell:1997ke}
  where the Clebsch-Gordan coefficients for a simply reducible direct
  product representation are given.  In the case at hand one simply
  considers an arbitrary representation in place of the direct product
  to arrive at our result.} and in our case one has the following
formula for the symmetrized link fields:
\twocol{
\begin{eqnarray}
\label{eq:gaugelink}
\lefteqn{\link^{\irrep^{PC}}_{\irrepi}(x)=} \\
& & \sum_{(i,j)}\frac{\sum_{\elem \in \Ogroup}
  \rep{U^{PC}}{(i,j)}{(m,n)}{\elem}
  \rep{\irrep}{\irrepi}{\irrepj}{\elem}^*}
{\left[\frac{\order{\Ogroup}}{\irrepdim_\irrep}
    \sum_{\elem \in \Ogroup}
    \rep{U^{PC}}{(m,n)}{(m,n)}{\elem}
    \rep{\irrep}{\irrepj}{\irrepj}{\elem}^*\right]^{\frac{1}{2}}}U^{PC}_{i,j}(x)
\ , \nonumber
\end{eqnarray}
}
{
\begin{equation}
\label{eq:gaugelink}
\link^{\irrep^{PC}}_{\irrepi}(x)= \sum_{(i,j)}\frac{\sum_{\elem \in \Ogroup}
  \rep{U^{PC}}{(i,j)}{(m,n)}{\elem}
  \rep{\irrep}{\irrepi}{\irrepj}{\elem}^*}
{\left[\frac{\order{\Ogroup}}{\irrepdim_\irrep}
    \sum_{\elem \in \Ogroup}
    \rep{U^{PC}}{(m,n)}{(m,n)}{\elem}
    \rep{\irrep}{\irrepj}{\irrepj}{\elem}^*\right]^{\frac{1}{2}}}U^{PC}_{i,j}(x)
\ ,
\end{equation}}
where $(m,n)$ and $\irrepj$ are chosen so the denominator does not
vanish and the accessible $\irrep^{PC}$ are taken from
Table~\ref{tab:upcreduction}.  Here
$\rep{\irrep}{\irrepi}{\irrepj}{\elem}$ is the matrix representation
for the irrep $\irrep$ of dimension $d_\irrep$ taken from
Table~\ref{tab:irrepmatrices}. Lowercase Greek denotes row indices of
the group $\Ogroup$, and the sum is taken over the six pairs $(i,j)$
listed above.  The $\link^{\irrep^{PC}}_{\irrepi}(x)$ transform as the
row $\irrepi$ of irrep $\irrep^{PC}$ and may be uniquely identified by
these latter parameters.  (Since each $\irrep^{PC}$ occurs only once
in Table~\ref{tab:upcreduction} there is no need, unlike in the spinor
case, to further identify a multiplicity for the irrep.)  Using the
explicit irrep generator matrices in Table~\ref{tab:irrepmatrices} to
construct the full matrix representations for each irrep, one can
evaluate Equation~(\ref{eq:gaugelink}) to produce the final
lattice-symmetrized gauge links.  The results are tabulated in
Table~\ref{tab:gaugelink}.  A diagram of a lattice-symmetrized gauge
link structure is shown in Figure~\ref{fig:Ut1mp}.

In Equation~(\ref{eq:pcadaptedu}), all the link structures with $PC\neq++$ are
suppressed by powers of
lattice spacing relative to $PC=++$.  Further factors of lattice spacing can
appear when Equation~(\ref{eq:pcadaptedu}) is used to form a gauge structure like
the one shown in Figure~\ref{fig:Ut1mp}.  Such overall factors of lattice
spacing will not be of direct relevance to our study of the mass spectrum, but
would be of greater interest for decay constants and matrix elements.

%R Original file named mesonopslatex.tex
\begin{table*}
\caption{\label{tab:gaugelink}Shown are the lattice-symmetrized gauge links,
  $\link^{\irrep^{PC}}_{\irrepi}(x)$, derived from
  Equation~(\ref{eq:gaugelink}) in terms of the $PC$-adapted gauge
  field structures $U^{PC}_{i,j}(x)$.  Each subtable corresponds to a
  different $PC$ combination and contains the lattice-symmetrized
  gauge link combinations along the top, denoted by
  $\irrep^{\irrepi}$, and the $PC$-adapted structures along the left,
  denoted by $U(i,j)$.  For irreps of dimension one the superscript
  denoting the row $\irrepi$ of the lattice-symmetrized link structure
  is suppressed.  The actual coefficient equals the sign of the entry
  times the \emph{square root} of the absolute value of the entry in
  the table.  Thus, for example, one has
  $\link^{T_1^{-+}}_{3}=\left(U^{-+}_{1,2}+U^{-+}_{1,3}+U^{-+}_{2,3}-U^{-+}_{2,1}\right)/2$,
  which in turn can be expanded via
  Equation~(\ref{eq:pcadaptedu}) to arrive at Figure~\ref{fig:Ut1mp}.}
\twocol{}{\begin{footnotesize}}
\begin{tabular}{ll}
$\begin{array}{|c|rrrrrr|} \hline \hline
PC=++ & A_{1} & E^{1} & E^{2} & T_{2}^{1} & T_{2}^{2} & T_{2}^{3}\\ \hline
U(1,2) & \ 1/6 & \ 1/4 & -1/12 & \ \ \ 0 & \ 1/2 & \ \ \ 0 \\ 
U(1,3) & \ 1/6 & \ 1/4 & -1/12 & \ \ \ 0 & -1/2 & \ \ \ 0 \\ 
U(2,3) & \ 1/6 & -1/4 & -1/12 & \ \ \ 0 & \ \ \ 0 & \ 1/2 \\ 
U(2,1) & \ 1/6 & -1/4 & -1/12 & \ \ \ 0 & \ \ \ 0 & -1/2 \\ 
U(3,1) & \ 1/6 & \ \ \ 0 & \ 1/3 & \ 1/2 & \ \ \ 0 & \ \ \ 0 \\ 
U(3,2) & \ 1/6 & \ \ \ 0 & \ 1/3 & -1/2 & \ \ \ 0 & \ \ \ 0 \\ \hline \hline
\end{array}$
&
$\begin{array}{|c|rrrrrr|} \hline \hline
PC=+- & A_{2} & E^{1} & E^{2} & T_{1}^{1} & T_{1}^{2} & T_{1}^{3}\\ \hline
U(1,2) & \ 1/6 & 1/12 & \ 1/4 & \ \; 1/2 & \ \ \ 0 & \ \ \ 0 \\ 
U(1,3) & -1/6 & -1/12 & -1/4 & \ \; 1/2 & \ \ \ 0 & \ \ \ 0 \\ 
U(2,3) & \ 1/6 & 1/12 & -1/4 & \ \ \ 0 & \ \; 1/2 & \ \ \ 0 \\ 
U(2,1) & -1/6 & -1/12 & \ 1/4 & \ \ \ 0 & \ \; 1/2 & \ \ \ 0 \\ 
U(3,1) & \ 1/6 & -1/3 & \ \ \ 0 & \ \ \ 0 & \ \ \ 0 & \ \; 1/2 \\ 
U(3,2) & -1/6 & \ 1/3 & \ \ \ 0 & \ \ \ 0 & \ \ \ 0 & \ \; 1/2 \\ \hline \hline
\end{array}$ \vspace{1ex}
\\ 
$\begin{array}{|c|rrrrrr|} \hline \hline
PC=-+ & T_{1}^{1} & T_{1}^{2} & T_{1}^{3} & T_{2}^{1} & T_{2}^{2} & T_{2}^{3}\\ \hline
U(1,2) & \ \ \ 0 & \ 1/4 & \ 1/4 & \ 1/4 & \ \ \ 0 & -1/4 \\ 
U(1,3) & \ \ \ 0 & -1/4 & \ 1/4 & \ 1/4 & \ \ \ 0 & \ 1/4 \\ 
U(2,3) & \ 1/4 & \ \ \ 0 & \ 1/4 & -1/4 & \ 1/4 & \ \ \ 0 \\ 
U(2,1) & \ 1/4 & \ \ \ 0 & -1/4 & \ 1/4 & \ 1/4 & \ \ \ 0 \\ 
U(3,1) & \ 1/4 & \ 1/4 & \ \ \ 0 & \ \ \ 0 & -1/4 & \ 1/4 \\ 
U(3,2) & -1/4 & \ 1/4 & \ \ \ 0 & \ \ \ 0 & \ 1/4 & \ 1/4 \\ \hline \hline
\end{array}$
&
$\begin{array}{|c|rrrrrr|} \hline \hline
PC=-- & T_{1}^{1} & T_{1}^{2} & T_{1}^{3} & T_{2}^{1} & T_{2}^{2} & T_{2}^{3}\\ \hline
U(1,2) & \ \ \ 0 & -1/4 & \ 1/4 & \ 1/4 & \ \ \ 0 & \ 1/4 \\ 
U(1,3) & \ \ \ 0 & -1/4 & -1/4 & -1/4 & \ \ \ 0 & \ 1/4 \\ 
U(2,3) & \ 1/4 & \ \ \ 0 & -1/4 & \ 1/4 & \ 1/4 & \ \ \ 0 \\ 
U(2,1) & -1/4 & \ \ \ 0 & -1/4 & \ 1/4 & -1/4 & \ \ \ 0 \\ 
U(3,1) & -1/4 & \ 1/4 & \ \ \ 0 & \ \ \ 0 & \ 1/4 & \ 1/4 \\ 
U(3,2) & -1/4 & -1/4 & \ \ \ 0 & \ \ \ 0 & \ 1/4 & -1/4 \\ \hline \hline
\end{array}$
\end{tabular}
\twocol{}{\end{footnotesize}}
\end{table*}

\subsection{Total representation reduction and operator construction}
\label{sec:total-repr-reduct}
Having classified the spinor and link components of our operators into
irreps of $\Ogroup^{PC}$, it remains to combine them and reduce all
possible direct product representations.  Parity and charge
conjugation for the product representations are given by
\begin{equation}
\begin{array}{ccc}
P=P_\diraclower P_\linklower & \hspace{1em} & C=C_\diraclower C_\linklower\ ,
\end{array}
\label{eq:3}
\end{equation}
where $\diraclower$ corresponds to the fermionic part and $\linklower$
corresponds to the gauge part.  Noting that the character of a direct
product representation like $\irrepdirac\otimes\irreplink$ satisfies
$\character^{(\irrepdirac\otimes\irreplink)}(\class)=\character^{(\irrepdirac)}(\class)\character^{(\irreplink)}(\class)$
and using Equation~(\ref{eq:2}) with $J=\irrepdirac\otimes\irreplink$
and the characters in Table~\ref{tab:ochartable}, one may reduce the
octahedral part of each direct product into irreps of $\Ogroup$ as
shown in Table~\ref{tab:crossreduction}.

\begin{table}
\caption{\label{tab:crossreduction}Reduction of product representations $\irrep_\diraclower\otimes\irrep_\linklower$ into irreps of $\Ogroup$.  Only irreps $A_1$ and $T_1$ are accessible for $\irrep_\diraclower$.  Reduction of other octahedral
direct products may be inferred from Table~\ref{tab:octahedralcg}.}
\twocol{\small}{}
$
\begin{array}{|c|ccccc|}
\hline
\otimes & A_1 & A_2 & E & T_1 & T_2 \\ \hline
A_1 & A_1 & A_2 & E & T_1 & T_2 \\
T_1 & T_1 & T_2 & T_1 \oplus T_2 & A_1 \oplus E \oplus T_1 \oplus T_2
& A_2 \oplus E \oplus T_1 \oplus T_2 \\
\hline
\end{array}
$
\end{table}

The results shown in Tables~\ref{tab:spinorreduction},
\ref{tab:upcreduction}, and \ref{tab:crossreduction} along with
Equation~(\ref{eq:3}) finally allow the $384$-dimensional
representation generated by the space spanned by $M_{j,k,a,b}(t)$ to
be reduced into irreps of $\Ogroup^{PC}$ as shown in
Table~\ref{tab:totalreduction}.  It is of note that while the local
quark operators can access only irreps of the form $A_1^{PC}$ and
$T_1^{PC}$ of $\Ogroup^{PC}$, the addition of extended gauge field
structure admits operators of every possible irrep.  In principle,
then, this set of operators can couple to all meson states.

\begin{table}
\caption{Reduction of the representation generated by the span of $M_{j,k,a,b}(t)$ into $\Ogroup^{PC}$ irreps.  The \mbox{384-dimensional} representation reduces into the sum of the irreps listed in this table with multiplicities shown.}
\label{tab:totalreduction}
$
\begin{array}{|c|rrrr|}
\hline
\irrep\backslash PC & ++ & +- & -+ & -- \\ \hline
A_1 & 4 & 4 & 6 & 2 \\
A_2 & 4 & 4 & 2 & 6 \\
E & 8 & 8 & 8 & 8 \\
T_1 & 12 & 12 & 10 & 14 \\
T_2 & 12 & 12 & 14 & 10 \\ \hline
\end{array}
$
\end{table}

The construction of the lattice-symmetrized operators themselves is
accomplished by combining the spinor and link operator components
$\dirac^{\irrep^{\irrepcopy PC}}_\irrepi$ and
$\link^{\irrep^{PC}}_\irrepi$ using the same formulae for parity and
charge conjugation given in Equation (\ref{eq:3}) and by combining the
irreps of $\Ogroup$ using Clebsch-Gordan (C-G) coefficients for the
reduction of a direct product of two irreps of $\Ogroup$,
$\irrepone\otimes\irreptwo$.  As suggested by
Table~\ref{tab:crossreduction}, the direct product representations for
irreps of $\Ogroup$ are simply reducible, allowing one to use the
formula
\twocol{
\begin{eqnarray}
\lefteqn{\left(\begin{array}{cc|c}
\irrepone & \irreptwo & \irrep \\
\irreponei & \irreptwoi & \irrepi
\end{array}
\right) =} \nonumber \\
&  & \frac{\sum_{\elem \in \Ogroup}
  \rep{\irrepone}{\irreponei}{\irreponej}{\elem}
  \rep{\irreptwo}{\irreptwoi}{\irreptwoj}{\elem}
  \rep{\irrep}{\irrepi}{\irrepj}{\elem}^*}
{\left[\frac{\order{\Ogroup}}{\irrepdim_\irrep}
    \sum_{\elem \in \Ogroup}
    \rep{\irrepone}{\irreponej}{\irreponej}{\elem}
    \rep{\irreptwo}{\irreptwoj}{\irreptwoj}{\elem}
    \rep{\irrep}{\irrepj}{\irrepj}{\elem}^*\right]^{\frac{1}{2}}}
\label{eq:octahedralcg}
\end{eqnarray}
}
{
\begin{equation}
\left(\begin{array}{cc|c}
\irrepone & \irreptwo & \irrep \\
\irreponei & \irreptwoi & \irrepi
\end{array}
\right)
= \frac{\sum_{\elem \in \Ogroup}
  \rep{\irrepone}{\irreponei}{\irreponej}{\elem}
  \rep{\irreptwo}{\irreptwoi}{\irreptwoj}{\elem}
  \rep{\irrep}{\irrepi}{\irrepj}{\elem}^*}
{\left[\frac{\order{\Ogroup}}{\irrepdim_\irrep}
    \sum_{\elem \in \Ogroup}
    \rep{\irrepone}{\irreponej}{\irreponej}{\elem}
    \rep{\irreptwo}{\irreptwoj}{\irreptwoj}{\elem}
    \rep{\irrep}{\irrepj}{\irrepj}{\elem}^*\right]^{\frac{1}{2}}}
\label{eq:octahedralcg}
\end{equation}
} to determine the C-G coefficients.  (See Ref.~\mbox{\cite[page
  74]{Cornwell:1997ke}}.)  Here $\irrepj$, $\irreponej$, and
$\irreptwoj$ are chosen so the denominator does not vanish.  Using the
irrep generators given in Table~\ref{tab:irrepmatrices}, these C-G
coefficients have been calculated and are given in
Table~\ref{tab:octahedralcg}.  The lattice-symmetrized operators,
$M^{\irrep^{PC}}_\irrepi(t)$, are finally \twocol{
\begin{eqnarray}
\lefteqn{
\mesonop_\irrepi^{\irrep^{PC},\irrepdirac^{\irrepcopy_\diraclower P_\diraclower C_\diraclower} , 
\irreplink^{P_\linklower C_\linklower}}(t) 
=}\label{eq:finalsymops}\\
& & {\displaystyle\sum_\vecx\sum_{\irrepdiraci, \irreplinki}}
\left(\begin{array}{cc|c}
\irrepdirac & \irreplink & \irrep \\
\irrepdiraci & \irreplinki & \irrepi
\end{array}
\right)
\spinorbar(x)\dirac^{\irrepdirac^{\irrepcopy_\diraclower P_\diraclower C_\diraclower}}_{\irrepdiraci}
\link^{\irreplink^{P_\linklower C_\linklower}}_{\irreplinki}(x)\spinor(x)
\;,\nonumber
\end{eqnarray}
}
{
\begin{equation}
\mesonop_\irrepi^{\irrep^{PC},\irrepdirac^{\irrepcopy_\diraclower P_\diraclower C_\diraclower}, 
\irreplink^{P_\linklower C_\linklower}}(t) 
={\displaystyle\sum_\vecx\sum_{\irrepdiraci, \irreplinki}}
\left(\begin{array}{cc|c}
\irrepdirac & \irreplink & \irrep \\
\irrepdiraci & \irreplinki & \irrepi
\end{array}
\right)
\spinorbar(x)\dirac^{\irrepdirac^{\irrepcopy_\diraclower P_\diraclower C_\diraclower}}_{\irrepdiraci}
\link^{\irreplink^{P_\linklower C_\linklower}}_{\irreplinki}(x)\spinor(x)
\ ,
\label{eq:finalsymops}
\end{equation}
}
where the allowed irreps for the spinor bilinear and link components
are determined by Tables~\ref{tab:spinorreduction}
and~\ref{tab:upcreduction}.  These fix $P$ and $C$ for the operator
while the irreps $\irrep$ of $\Ogroup$ are those allowed by the C-G
series in Table~\ref{tab:crossreduction}.  Dirac indices on $\dirac$
and color indices on $\link$ and both indices on the spinors have been
suppressed.  We note that each operator is thus uniquely identified by
its irrep $\irrep$, its row $\irrepi$, and the direct product from
which it originates, $\irrepdirac^{\irrepcopy_\diraclower
  P_\diraclower C_\diraclower}\otimes \irreplink^{P_\linklower
  C_\linklower}$.

\begin{table*}
\caption{\label{tab:octahedralcg}Octahedral group Clebsch-Gordan coefficients derived from Equation (\ref{eq:octahedralcg}).  The actual coefficient equals the sign of the entry times the \emph{square root} of its absolute value. Superscripts on irreps denote the row.  If the irrep is one-dimensional the superscript is suppressed.  Since the C-G coefficients for $\irrepone\otimes\irreptwo$ are the same as for $\irreptwo\otimes\irrepone$ only one combination is listed.}
\twocol{}{\begin{footnotesize}}
\begin{tabular}{l@{\hspace{-1ex}}l}
\begin{tabular}{l}
\begin{tabular}{ll}

\begin{tabular}{l}
$\begin{array}{|c|r|} \hline \hline
A_{1} \otimes A_{1} & A_{1}\\ \hline
A_{1}A_{1}  & \ \ \ 1 \\ \hline \hline
\end{array}$ \vspace{1ex}
\\
$\begin{array}{|c|r|} \hline \hline
A_{1} \otimes A_{2} & A_{2}\\ \hline
A_{1}A_{2}  & \ \ \ 1 \\ \hline \hline
\end{array}$ \vspace{1ex}
\\
$\begin{array}{|c|r|} \hline \hline
A_{2} \otimes A_{2} & A_{1}\\ \hline
A_{2}A_{2}  & \ \ \ 1 \\ \hline \hline
\end{array}$
\end{tabular}

&

\begin{tabular}{l}
$\begin{array}{|c|rr|} \hline \hline
A_{1} \otimes E & E^{1} & E^{2}\\ \hline
A_{1}E^{1}  & \ \ \ 1 & \ \ \ 0 \\ 
A_{1}E^{2}  & \ \ \ 0 & \ \ \ 1 \\ \hline \hline
\end{array}$ \vspace{3.5ex}
\\
$\begin{array}{|c|rr|} \hline \hline
A_{2} \otimes E & E^{1} & E^{2}\\ \hline
A_{2}E^{1}  & \ \ \ 0 & \ \ \ 1 \\ 
A_{2}E^{2}  & \ \ -1 & \ \ \ 0 \\ \hline \hline
\end{array}$
\end{tabular}

\end{tabular} \vspace{1ex}
\\
\begin{tabular}{ll}
$\begin{array}{|c|rrr|} \hline \hline
A_{1} \otimes T_{1} & T_{1}^{1} & T_{1}^{2} & T_{1}^{3}\\ \hline
A_{1}T_{1}^{1}  & \ \ \ 1 & \ \ \ 0 & \ \ \ 0 \\ 
A_{1}T_{1}^{2}  & \ \ \ 0 & \ \ \ 1 & \ \ \ 0 \\ 
A_{1}T_{1}^{3}  & \ \ \ 0 & \ \ \ 0 & \ \ \ 1 \\ \hline \hline
\end{array}$ \vspace{1ex}
&
$\begin{array}{|c|rrr|} \hline \hline
A_{1} \otimes T_{2} & T_{2}^{1} & T_{2}^{2} & T_{2}^{3}\\ \hline
A_{1}T_{2}^{1}  & \ \ \ 1 & \ \ \ 0 & \ \ \ 0 \\ 
A_{1}T_{2}^{2}  & \ \ \ 0 & \ \ \ 1 & \ \ \ 0 \\ 
A_{1}T_{2}^{3}  & \ \ \ 0 & \ \ \ 0 & \ \ \ 1 \\ \hline \hline
\end{array}$
\end{tabular} \vspace{1ex}
\\
\begin{tabular}{ll}
$\begin{array}{|c|rrr|} \hline \hline
A_{2} \otimes T_{1} & T_{2}^{1} & T_{2}^{2} & T_{2}^{3}\\ \hline
A_{2}T_{1}^{1}  & \ \ \ 0 & \ \ \ 1 & \ \ \ 0 \\ 
A_{2}T_{1}^{2}  & \ \ \ 0 & \ \ \ 0 & \ \ \ 1 \\ 
A_{2}T_{1}^{3}  & \ \ \ 1 & \ \ \ 0 & \ \ \ 0 \\ \hline \hline
\end{array}$ \vspace{1ex}
&
$\begin{array}{|c|rrr|} \hline \hline
A_{2} \otimes T_{2} & T_{1}^{1} & T_{1}^{2} & T_{1}^{3}\\ \hline
A_{2}T_{2}^{1}  & \ \ \ 0 & \ \ \ 0 & \ \ \ 1 \\ 
A_{2}T_{2}^{2}  & \ \ \ 1 & \ \ \ 0 & \ \ \ 0 \\ 
A_{2}T_{2}^{3}  & \ \ \ 0 & \ \ \ 1 & \ \ \ 0 \\ \hline \hline
\end{array}$
\end{tabular} \vspace{1ex}
\\
$\begin{array}{|c|rrrr|} \hline \hline
E \otimes E & A_{1} & A_{2} & E^{1} & E^{2}\\ \hline
E^{1}E^{1}  & \ 1/2 & \ \ \ 0 & \ \ \ 0 & \ 1/2 \\ 
E^{1}E^{2}  & \ \ \ 0 & \ 1/2 & \ 1/2 & \ \ \ 0 \\ 
E^{2}E^{1}  & \ \ \ 0 & -1/2 & \ 1/2 & \ \ \ 0 \\ 
E^{2}E^{2}  & \ 1/2 & \ \ \ 0 & \ \ \ 0 & -1/2 \\ \hline \hline
\end{array}$ \vspace{1ex}
\\
$\begin{array}{|c|rrrrrr|} \hline \hline
E \otimes T_{1} & T_{1}^{1} & T_{1}^{2} & T_{1}^{3} & T_{2}^{1} & T_{2}^{2} & T_{2}^{3}\\ \hline
E^{1}T_{1}^{1}  & \ 3/4 & \ \ \ 0 & \ \ \ 0 & \ \ \ 0 & \ 1/4 & \ \ \ 0 \\ 
E^{1}T_{1}^{2}  & \ \ \ 0 & -3/4 & \ \ \ 0 & \ \ \ 0 & \ \ \ 0 & \ 1/4 \\ 
E^{1}T_{1}^{3}  & \ \ \ 0 & \ \ \ 0 & \ \ \ 0 & \ \ -1 & \ \ \ 0 & \ \ \ 0 \\ 
E^{2}T_{1}^{1}  & -1/4 & \ \ \ 0 & \ \ \ 0 & \ \ \ 0 & \ 3/4 & \ \ \ 0 \\ 
E^{2}T_{1}^{2}  & \ \ \ 0 & -1/4 & \ \ \ 0 & \ \ \ 0 & \ \ \ 0 & -3/4 \\ 
E^{2}T_{1}^{3}  & \ \ \ 0 & \ \ \ 0 & \ \ \ 1 & \ \ \ 0 & \ \ \ 0 & \ \ \ 0 \\ \hline \hline
\end{array}$ \vspace{1ex}
\\
$\begin{array}{|c|rrrrrr|} \hline \hline
E \otimes T_{2} & T_{1}^{1} & T_{1}^{2} & T_{1}^{3} & T_{2}^{1} & T_{2}^{2} & T_{2}^{3}\\ \hline
E^{1}T_{2}^{1}  & \ \ \ 0 & \ \ \ 0 & \ \ \ 1 & \ \ \ 0 & \ \ \ 0 & \ \ \ 0 \\ 
E^{1}T_{2}^{2}  & -1/4 & \ \ \ 0 & \ \ \ 0 & \ \ \ 0 & \ 3/4 & \ \ \ 0 \\ 
E^{1}T_{2}^{3}  & \ \ \ 0 & -1/4 & \ \ \ 0 & \ \ \ 0 & \ \ \ 0 & -3/4 \\ 
E^{2}T_{2}^{1}  & \ \ \ 0 & \ \ \ 0 & \ \ \ 0 & \ \ \ 1 & \ \ \ 0 & \ \ \ 0 \\ 
E^{2}T_{2}^{2}  & -3/4 & \ \ \ 0 & \ \ \ 0 & \ \ \ 0 & -1/4 & \ \ \ 0 \\ 
E^{2}T_{2}^{3}  & \ \ \ 0 & \ 3/4 & \ \ \ 0 & \ \ \ 0 & \ \ \ 0 & -1/4 \\ \hline \hline
\end{array}$
\end{tabular} &
\begin{tabular}{l}
$\begin{array}{|c|rrrrrrrrr|} \hline \hline
T_{1} \otimes T_{1} & A_{1} & E^{1} & E^{2} & T_{1}^{1} & T_{1}^{2} & T_{1}^{3} & T_{2}^{1} & T_{2}^{2} & T_{2}^{3}\\ \hline
T_{1}^{1}T_{1}^{1}  & \ 1/3 & \ 1/2 & -1/6 & \ \ \ 0 & \ \ \ 0 & \ \ \ 0 & \ \ \ 0 & \ \ \ 0 & \ \ \ 0 \\ 
T_{1}^{1}T_{1}^{2}  & \ \ \ 0 & \ \ \ 0 & \ \ \ 0 & \ \ \ 0 & \ \ \ 0 & \ 1/2 & \ 1/2 & \ \ \ 0 & \ \ \ 0 \\ 
T_{1}^{1}T_{1}^{3}  & \ \ \ 0 & \ \ \ 0 & \ \ \ 0 & \ \ \ 0 & -1/2 & \ \ \ 0 & \ \ \ 0 & \ \ \ 0 & \ 1/2 \\ 
T_{1}^{2}T_{1}^{1}  & \ \ \ 0 & \ \ \ 0 & \ \ \ 0 & \ \ \ 0 & \ \ \ 0 & -1/2 & \ 1/2 & \ \ \ 0 & \ \ \ 0 \\ 
T_{1}^{2}T_{1}^{2}  & \ 1/3 & -1/2 & -1/6 & \ \ \ 0 & \ \ \ 0 & \ \ \ 0 & \ \ \ 0 & \ \ \ 0 & \ \ \ 0 \\ 
T_{1}^{2}T_{1}^{3}  & \ \ \ 0 & \ \ \ 0 & \ \ \ 0 & \ 1/2 & \ \ \ 0 & \ \ \ 0 & \ \ \ 0 & \ 1/2 & \ \ \ 0 \\ 
T_{1}^{3}T_{1}^{1}  & \ \ \ 0 & \ \ \ 0 & \ \ \ 0 & \ \ \ 0 & \ 1/2 & \ \ \ 0 & \ \ \ 0 & \ \ \ 0 & \ 1/2 \\ 
T_{1}^{3}T_{1}^{2}  & \ \ \ 0 & \ \ \ 0 & \ \ \ 0 & -1/2 & \ \ \ 0 & \ \ \ 0 & \ \ \ 0 & \ 1/2 & \ \ \ 0 \\ 
T_{1}^{3}T_{1}^{3}  & \ 1/3 & \ \ \ 0 & \ 2/3 & \ \ \ 0 & \ \ \ 0 & \ \ \ 0 & \ \ \ 0 & \ \ \ 0 & \ \ \ 0 \\ \hline \hline
\end{array}$\vspace{12ex}
\\
$\begin{array}{|c|rrrrrrrrr|} \hline \hline
T_{1} \otimes T_{2} & A_{2} & E^{1} & E^{2} & T_{1}^{1} & T_{1}^{2} & T_{1}^{3} & T_{2}^{1} & T_{2}^{2} & T_{2}^{3}\\ \hline
T_{1}^{1}T_{2}^{1}  & \ \ \ 0 & \ \ \ 0 & \ \ \ 0 & \ \ \ 0 & \ 1/2 & \ \ \ 0 & \ \ \ 0 & \ \ \ 0 & \ 1/2 \\ 
T_{1}^{1}T_{2}^{2}  & \ 1/3 & \ 1/6 & \ 1/2 & \ \ \ 0 & \ \ \ 0 & \ \ \ 0 & \ \ \ 0 & \ \ \ 0 & \ \ \ 0 \\ 
T_{1}^{1}T_{2}^{3}  & \ \ \ 0 & \ \ \ 0 & \ \ \ 0 & \ \ \ 0 & \ \ \ 0 & \ 1/2 & -1/2 & \ \ \ 0 & \ \ \ 0 \\ 
T_{1}^{2}T_{2}^{1}  & \ \ \ 0 & \ \ \ 0 & \ \ \ 0 & \ 1/2 & \ \ \ 0 & \ \ \ 0 & \ \ \ 0 & -1/2 & \ \ \ 0 \\ 
T_{1}^{2}T_{2}^{2}  & \ \ \ 0 & \ \ \ 0 & \ \ \ 0 & \ \ \ 0 & \ \ \ 0 & \ 1/2 & \ 1/2 & \ \ \ 0 & \ \ \ 0 \\ 
T_{1}^{2}T_{2}^{3}  & \ 1/3 & \ 1/6 & -1/2 & \ \ \ 0 & \ \ \ 0 & \ \ \ 0 & \ \ \ 0 & \ \ \ 0 & \ \ \ 0 \\ 
T_{1}^{3}T_{2}^{1}  & \ 1/3 & -2/3 & \ \ \ 0 & \ \ \ 0 & \ \ \ 0 & \ \ \ 0 & \ \ \ 0 & \ \ \ 0 & \ \ \ 0 \\ 
T_{1}^{3}T_{2}^{2}  & \ \ \ 0 & \ \ \ 0 & \ \ \ 0 & \ \ \ 0 & \ 1/2 & \ \ \ 0 & \ \ \ 0 & \ \ \ 0 & -1/2 \\ 
T_{1}^{3}T_{2}^{3}  & \ \ \ 0 & \ \ \ 0 & \ \ \ 0 & \ 1/2 & \ \ \ 0 & \ \ \ 0 & \ \ \ 0 & \ 1/2 & \ \ \ 0 \\ \hline \hline
\end{array}$\vspace{12ex}
\\
$\begin{array}{|c|rrrrrrrrr|} \hline \hline
T_{2} \otimes T_{2} & A_{1} & E^{1} & E^{2} & T_{1}^{1} & T_{1}^{2} & T_{1}^{3} & T_{2}^{1} & T_{2}^{2} & T_{2}^{3}\\ \hline
T_{2}^{1}T_{2}^{1}  & \ 1/3 & \ \ \ 0 & \ 2/3 & \ \ \ 0 & \ \ \ 0 & \ \ \ 0 & \ \ \ 0 & \ \ \ 0 & \ \ \ 0 \\ 
T_{2}^{1}T_{2}^{2}  & \ \ \ 0 & \ \ \ 0 & \ \ \ 0 & \ \ \ 0 & \ 1/2 & \ \ \ 0 & \ \ \ 0 & \ \ \ 0 & \ 1/2 \\ 
T_{2}^{1}T_{2}^{3}  & \ \ \ 0 & \ \ \ 0 & \ \ \ 0 & -1/2 & \ \ \ 0 & \ \ \ 0 & \ \ \ 0 & \ 1/2 & \ \ \ 0 \\ 
T_{2}^{2}T_{2}^{1}  & \ \ \ 0 & \ \ \ 0 & \ \ \ 0 & \ \ \ 0 & -1/2 & \ \ \ 0 & \ \ \ 0 & \ \ \ 0 & \ 1/2 \\ 
T_{2}^{2}T_{2}^{2}  & \ 1/3 & \ 1/2 & -1/6 & \ \ \ 0 & \ \ \ 0 & \ \ \ 0 & \ \ \ 0 & \ \ \ 0 & \ \ \ 0 \\ 
T_{2}^{2}T_{2}^{3}  & \ \ \ 0 & \ \ \ 0 & \ \ \ 0 & \ \ \ 0 & \ \ \ 0 & \ 1/2 & \ 1/2 & \ \ \ 0 & \ \ \ 0 \\ 
T_{2}^{3}T_{2}^{1}  & \ \ \ 0 & \ \ \ 0 & \ \ \ 0 & \ 1/2 & \ \ \ 0 & \ \ \ 0 & \ \ \ 0 & \ 1/2 & \ \ \ 0 \\ 
T_{2}^{3}T_{2}^{2}  & \ \ \ 0 & \ \ \ 0 & \ \ \ 0 & \ \ \ 0 & \ \ \ 0 & -1/2 & \ 1/2 & \ \ \ 0 & \ \ \ 0 \\ 
T_{2}^{3}T_{2}^{3}  & \ 1/3 & -1/2 & -1/6 & \ \ \ 0 & \ \ \ 0 & \ \ \ 0 & \ \ \ 0 & \ \ \ 0 & \ \ \ 0 \\ \hline \hline
\end{array}$
\end{tabular}
\end{tabular}
\twocol{}{\end{footnotesize}}
\end{table*}

%%% Local Variables: 
%%% mode: latex
%%% TeX-master: "tmmeson"
%%% End: 

\section{\label{sec:twistedmass}Twisted mass}

It is a feature of twisted mass lattice QCD (tmLQCD) that,
at maximal twist, $O(a)$ errors are absent from physical
observables~\cite{Frezzotti:2000nk,Frezzotti:2003ni}.  Critical slow-down is
softened in tmLQCD and thus it permits simulations with light
quark masses.  Unlike theories that require the tuning of separate parameters for the
improvement of each
individual operator, tmLQCD requires the tuning of only one, the
standard mass parameter $m_0$.  At maximal twist, information about the
physical quark mass is given by the twisted mass parameter $\mu_0$.
However, the theory modifies the parity symmetry of QCD.  The implication
of this for meson correlators is examined in this section.

The tmLQCD fermion action is
\begin{equation}\label{tmLQCDaction1}
S_F = a^4\sum_{q=u,d}\sum_{x,y}\bar q(x)S_q^{-1}(x,y)q(y) \;,
\end{equation}
where
\twocol{
\begin{eqnarray}
\lefteqn{\sum_yS_q^{-1}(x,y)q(y)} \nonumber \\
 &=& \frac{1}{2a}\sum_\mu\gamma_\mu\left[U_\mu(x)
     q(x+a\hat\mu)-U^\dagger_\mu(x-a\hat\mu)q(x-a\hat\mu)\right] \nonumber \\
 &-& \frac{1}{2a}\sum_\mu\left[U_\mu(x)q(x+a\hat\mu)
     +U^\dagger_\mu(x-a\hat\mu)q(x-a\hat\mu)-2q(x)\right] \nonumber \\
 &+& \left[m_0 + i\gamma_5\mu_q\right]q(x) \label{tmLQCDaction2}
\end{eqnarray}
}
{
\begin{eqnarray}
\sum_yS_q^{-1}(x,y)q(y) &=& \frac{1}{2a}\sum_\mu\gamma_\mu\left[U_\mu(x)
     q(x+a\hat\mu)-U^\dagger_\mu(x-a\hat\mu)q(x-a\hat\mu)\right] \nonumber \\
 &-& \frac{1}{2a}\sum_\mu\left[U_\mu(x)q(x+a\hat\mu)
     +U^\dagger_\mu(x-a\hat\mu)q(x-a\hat\mu)-2q(x)\right] \nonumber \\
 &+& \left[m_0 + i\gamma_5\mu_q\right]q(x) \label{tmLQCDaction2}
\end{eqnarray}
}
and
\begin{equation}
\mu_u=-\mu_d\equiv \mu_0
\end{equation}
is the twisted mass parameter.
The tmLQCD action of Equations (\ref{tmLQCDaction1}-\ref{tmLQCDaction2})
is displayed in the ``twisted basis''.  Fermion propagators are computed
in that basis and then converted to the ``physical basis'' via
\begin{eqnarray}
(S_u)_{\rm physical} &=& e^{i\gamma_5\pi/4}(S_u)_{\rm twisted}e^{i\gamma_5\pi/4} \;, \\
(S_d)_{\rm physical} &=& e^{-i\gamma_5\pi/4}(S_d)_{\rm twisted}e^{-i\gamma_5\pi/4} \;.
\end{eqnarray}
Except for Equations (\ref{tmLQCDaction1}-\ref{tmLQCDaction2}) and
(\ref{tuningeq}), all discussions in this work are in the physical basis
at maximal twist.

While parity $P$ defined in standard fashion by Equation (\ref{PandC})
is not a symmetry of the tmLQCD action, flavor-parity
\begin{equation}
\tilde{P} \equiv F_2P
\end{equation}
which combines the
parity operation with a 180$^\circ$ rotation about the second isospin axis,
\begin{equation}
\begin{array}{ccc}
\mathcal{F}_2u\mathcal{F}_2^\dagger = -d, &
\hspace{1em} &
\mathcal{F}_2d\mathcal{F}_2^\dagger = u, \\
\mathcal{F}_2\overline{u}\mathcal{F}_2^\dagger = -\overline{d}, &
&
\mathcal{F}_2\overline{d}\mathcal{F}_2^\dagger = \overline{u},
\end{array}
\end{equation}
is respected~\cite{Frezzotti:2000nk}.
In our notation, $\mathcal{F}_2$ is defined by
\begin{equation}
\mathcal{F}_2 \equiv e^{i\pi\tau_2}
\end{equation}
where $\tau_2$ is an isospin operator, defined in Ref.~\cite{Basak:2005aq},
whose action on quark fields can be inferred from the commutation relations in
Equations (31) and (32) found therein.

Notice the relations involving reflections,
\begin{equation}
S_u^{-1}(0,y;U) = \gamma_i\gamma_5S^{-1}_d(0,y_I;U_I)\gamma_5\gamma_i\;,
\end{equation}
where the subscript $I$ refers to the lattice coordinates and links
after inversion in the $i$th direction.
Applying this relation in
all three spatial directions and a consideration of the inverse leads
to a flavor-parity relation
\begin{equation}\label{flavorparity}
S_u(x,y;U) = \gamma_4S_d(x_P,y_P;U_P)\gamma_4\;,
\end{equation}
where $x_P$ is defined by Equation (\ref{xP}) and $U_P$ represents the links
after inversion under spatial parity.
The tmLQCD action also preserves charge conjugation,
\begin{equation}
S_u(U) = CS_u^T(U^*)C^\dagger
\end{equation}
and has a Hermiticity relation,
\begin{equation}\label{Hermiticity}
S_u(U) = \gamma_5S_d^\dagger(U)\gamma_5 \;.
\end{equation}

In our simulations we consider charged-meson two-point correlators,
\twocol{
\begin{eqnarray}
\lefteqn{C_{AB}(t) =} \\
& & \left<\sum_{\stackrel{\scriptstyle \vec{x},{\rm spins}}{{\rm colors}}}
            \bar u(0)\gamma_4(\dirac^A\link^A(0))^\dagger\gamma_4d(0)
            \bar d(x)\dirac^B\link^B(x)u(x)\right> \nonumber
\end{eqnarray}
}
{
\begin{equation}
C_{AB}(t) = \left<\sum_{\vec x}\sum_{\rm spins}\sum_{\rm colors}
            \bar u(0)\gamma_4(\dirac^A\link^A(0))^\dagger\gamma_4d(0)
            \bar d(x)\dirac^B\link^B(x)u(x)\right>
\end{equation}
}
and neutral-meson two-point correlators with disconnected contractions omitted,
\twocol{
\begin{eqnarray}
\lefteqn{N_{AB}(t) =} \nonumber \\
& &  \left<\sum_{\stackrel{\scriptstyle \vec{x},{\rm spins}}{{\rm colors}}}
            \bar u(0)\gamma_4(\dirac^A\link^A(0))^\dagger\gamma_4u(0)
            \bar u(x)\dirac^B\link^B(x)u(x)\right> \nonumber \\
& &        + (u\leftrightarrow d)\;,
\end{eqnarray}
}
{
\begin{equation}
N_{AB}(t) = \left<\sum_{\vec x}\sum_{\rm spins}\sum_{\rm colors}
            \bar u(0)\gamma_4(\dirac^A\link^A(0))^\dagger\gamma_4u(0)
            \bar u(x)\dirac^B\link^B(x)u(x)\right> + (u\leftrightarrow d)
\end{equation}
}
where each of $\link^A$ and $\link^B$ is one of the symmetrized link
fields $\link^{\irrep^{PC}}_{\irrepi}$ defined in
Equation~(\ref{eq:gaugelink}), and each of $\dirac^I$ and $\dirac^J$ is
one of the 16 unique products of Dirac matrices
$\dirac^{\irrep^{\irrepcopy PC}}_\irrepi$ defined in
Table~\ref{tab:spinorreduction}.
Choosing $\dirac^A$ to be Hermitian or anti-Hermitian and
evaluating the (connected) contractions, a few lines of algebra leads
to valuable conclusions:
\begin{itemize}
\item In the configuration average, $C_{AA}(t)$ is real for \mbox{tmLQCD} fermions.
\item In the configuration average, $N_{AA}(t)$ is real for \mbox{tmLQCD} fermions.
\end{itemize}
Note that the imaginary part of $N_{AA}(t)$ does not vanish if the
flavor symmetrization, $(u\leftrightarrow d)$, is omitted.
In practice, correlators in the charged case are averaged as well but for
improved statistics.

We are interested in correlators $C'_{IJ}(t)$ and $N'_{IJ}(t)$ between
the fully-symmetrized source and sink meson operators $I$ and $J$ of
the form $M^{\irrep^{PC}}_\irrepi$ given by
Equation~(\ref{eq:finalsymops}).  These correlators involve C-G
superpositions of $C_{AB}(t)$ and $N_{AB}(t)$ respectively and the
above results hold for $C'_{II}(t)$ and $N'_{II}(t)$ as well.

It is necessary to consider what effect the use of a twisted mass action will
have on the group theory discussion from Section~\ref{sec:latticeops}.
The distinction between respecting $\tilde P$ and not $P$ is clarified
by using the three symmetry relations,
Equations (\ref{flavorparity}-\ref{Hermiticity}), to evaluate two-point
correlators of operators of given parity $P$.
Table~\ref{tab:twopoint} shows the resulting orthogonalities among
various creation/annihilation operators.  Note that charged mesons,
containing $\bar ud$ or $\bar du$ creation operators, have different
relations from neutral mesons, containing $\bar uu$ or $\bar dd$
creation operators.

\begin{table*}
\caption{\label{tab:twopoint}Orthogonality relations among creation/annihilation operators in
  tmLQCD.  This table applies to any fixed \mbox{$\Lambda\in\{A_1,A_2,E,T_1,T_2\}$}.
Superscripts denote $P$ (not $\tilde P$) and $C$ for the operators $I$ and $J$.
The charged mesons are only eigenstates of $G$-parity with eigenvalue
$G=C(-1)^I=-C$, but we abuse the notation here and elsewhere in the
paper.}
\begin{center}
\begin{ruledtabular}
\begin{tabular}{ccrr}
$I$ & $J$ & \multicolumn{1}{c}{charged meson relations}
 & \multicolumn{1}{c}{neutral meson relations} \\
\hline
$\Lambda^{++}$ & $\Lambda^{++}$ &
$\left<\gamma_4I^\dagger\gamma_4S_uJS_d\right> - (u\leftrightarrow d) = 0$ &
$\left<\gamma_4I^\dagger\gamma_4S_uJS_u\right> - (u\leftrightarrow d) = 0$ \\
$\Lambda^{++}$ & $\Lambda^{+-}$ &
$\left<\gamma_4I^\dagger\gamma_4S_uJS_d\right> = 0$ &
$\left<\gamma_4I^\dagger\gamma_4S_uJS_u\right> = 0$ \\
$\Lambda^{++}$ & $\Lambda^{-+}$ &
$\left<\gamma_4I^\dagger\gamma_4S_uJS_d\right> = 0$ &
$\left<\gamma_4I^\dagger\gamma_4S_uJS_u\right> + (u\leftrightarrow d) = 0$ \\
$\Lambda^{++}$ & $\Lambda^{--}$ &
$\left<\gamma_4I^\dagger\gamma_4S_uJS_d\right> + (u\leftrightarrow d) = 0$ &
$\left<\gamma_4I^\dagger\gamma_4S_uJS_u\right> = 0$ \\
\hline
$\Lambda^{+-}$ & $\Lambda^{++}$ &
$\left<\gamma_4I^\dagger\gamma_4S_uJS_d\right> = 0$ &
$\left<\gamma_4I^\dagger\gamma_4S_uJS_u\right> = 0$ \\
$\Lambda^{+-}$ & $\Lambda^{+-}$ &
$\left<\gamma_4I^\dagger\gamma_4S_uJS_d\right> - (u\leftrightarrow d) = 0$ &
$\left<\gamma_4I^\dagger\gamma_4S_uJS_u\right> - (u\leftrightarrow d) = 0$ \\
$\Lambda^{+-}$ & $\Lambda^{-+}$ &
$\left<\gamma_4I^\dagger\gamma_4S_uJS_d\right> + (u\leftrightarrow d) = 0$ &
$\left<\gamma_4I^\dagger\gamma_4S_uJS_u\right> = 0$ \\
$\Lambda^{+-}$ & $\Lambda^{--}$ &
$\left<\gamma_4I^\dagger\gamma_4S_uJS_d\right> = 0$ &
$\left<\gamma_4I^\dagger\gamma_4S_uJS_u\right> + (u\leftrightarrow d) = 0$ \\
\hline
$\Lambda^{-+}$ & $\Lambda^{++}$ &
$\left<\gamma_4I^\dagger\gamma_4S_uJS_d\right> = 0$ &
$\left<\gamma_4I^\dagger\gamma_4S_uJS_u\right> + (u\leftrightarrow d) = 0$ \\
$\Lambda^{-+}$ & $\Lambda^{+-}$ &
$\left<\gamma_4I^\dagger\gamma_4S_uJS_d\right> + (u\leftrightarrow d) = 0$ &
$\left<\gamma_4I^\dagger\gamma_4S_uJS_u\right> = 0$ \\
$\Lambda^{-+}$ & $\Lambda^{-+}$ &
$\left<\gamma_4I^\dagger\gamma_4S_uJS_d\right> - (u\leftrightarrow d) = 0$ &
$\left<\gamma_4I^\dagger\gamma_4S_uJS_u\right> - (u\leftrightarrow d) = 0$ \\
$\Lambda^{-+}$ & $\Lambda^{--}$ &
$\left<\gamma_4I^\dagger\gamma_4S_uJS_d\right> = 0$ &
$\left<\gamma_4I^\dagger\gamma_4S_uJS_u\right> = 0$ \\
\hline
$\Lambda^{--}$ & $\Lambda^{++}$ &
$\left<\gamma_4I^\dagger\gamma_4S_uJS_d\right> + (u\leftrightarrow d) = 0$ &
$\left<\gamma_4I^\dagger\gamma_4S_uJS_u\right> = 0$ \\
$\Lambda^{--}$ & $\Lambda^{+-}$ &
$\left<\gamma_4I^\dagger\gamma_4S_uJS_d\right> = 0$ &
$\left<\gamma_4I^\dagger\gamma_4S_uJS_u\right> + (u\leftrightarrow d) = 0$ \\
$\Lambda^{--}$ & $\Lambda^{-+}$ &
$\left<\gamma_4I^\dagger\gamma_4S_uJS_d\right> = 0$ &
$\left<\gamma_4I^\dagger\gamma_4S_uJS_u\right> = 0$ \\
$\Lambda^{--}$ & $\Lambda^{--}$ &
$\left<\gamma_4I^\dagger\gamma_4S_uJS_d\right> - (u\leftrightarrow d) = 0$ &
$\left<\gamma_4I^\dagger\gamma_4S_uJS_u\right> - (u\leftrightarrow d) = 0$
\end{tabular}
\end{ruledtabular}
\end{center}
\end{table*}

In Table~\ref{tab:twopoint}, the charged meson entries that vanish
{\em without} a flavor interchange
$(u\leftrightarrow d)$ are direct consequences of $\tilde PC$ symmetry.
This is equivalent to the physical $PG$ symmetry since $G\equiv F_2C$.
These lines in the table
prove that $PG=-1$ states (i.e.\ $\Lambda^{++}$ and $\Lambda^{--}$) are
orthogonal to $PG=+1$ states (i.e.\ $\Lambda^{+-}$ and $\Lambda^{-+}$).
For example, because the physical pion is an eigenstate of
$PG$ with eigenvalue +1, we know that no charged pion signal can appear within
$A_1^{++}$ nor within $A_1^{--}$.

Unfortunately, the charged physical pion will in general couple to
both the $A_1^{+-}$ and $A_1^{-+}$ channels.  This is because it is an
eigenstate of $P$ and $G$ separately, whereas tmLQCD does not respect
these as good quantum numbers.  Instead, tmLQCD respects $\tilde P$
and $C$, but the charged pion is not an eigenstate of those.  Notice
that the lines in Table~\ref{tab:twopoint} that involve
``$+(u\leftrightarrow d)$'' merely show that unphysical states of
opposite flavor-parity are orthogonal, i.e.\
\twocol{
\begin{eqnarray}
\lefteqn{\left<\gamma_4I^\dagger\gamma_4S_uJS_d\right> + (u\leftrightarrow d)} \nonumber\\
&=& \left<\bar d\gamma_4I^\dagger\gamma_4u\bar uJd\right>
  + \left<\bar u\gamma_4I^\dagger\gamma_4d\bar dJu\right> \nonumber \\
&=& \left<\left(\bar d\gamma_4I^\dagger\gamma_4u+\bar u\gamma_4I^\dagger
    \gamma_4d\right)\left(\bar uJd+\bar dJu\right)\right>\; .
\end{eqnarray}
}
{
\begin{eqnarray}
\left<\gamma_4I^\dagger\gamma_4S_uJS_d\right> + (u\leftrightarrow d)
&=& \left<\bar d\gamma_4I^\dagger\gamma_4u\bar uJd\right>
  + \left<\bar u\gamma_4I^\dagger\gamma_4d\bar dJu\right> \nonumber \\
&=& \left<\left(\bar d\gamma_4I^\dagger\gamma_4u+\bar u\gamma_4I^\dagger
    \gamma_4d\right)\left(\bar uJd+\bar dJu\right)\right>\; .
\end{eqnarray}
}
Each factor in parentheses has definite flavor-parity, so finding
that this matrix element vanishes (as was found in some entries in the
table) means that the flavor-parities are orthogonal.  Similarly, the
charged physical mesons in $A_1^{++}$ and $A_1^{--}$ cannot be
separated either.

In Table~\ref{tab:twopoint}, the neutral meson entries that vanish
{\em without} using $(u\leftrightarrow d)$
are direct consequences of $C$ symmetry.  These lines prove that $C=+1$ states
(i.e.\ $\Lambda^{++}$ and $\Lambda^{-+}$) are orthogonal to $C=-1$ states
(i.e.\ $\Lambda^{+-}$ and $\Lambda^{--}$).
For example, because the physical neutral pion is an eigenstate of 
$C$ with eigenvalue $+1$, we know that no neutral pion signal can appear within
$A_1^{+-}$ nor within $A_1^{--}$.

In contrast to the case of the charged pion, the physical neutral pion
{\em is} an eigenstate of $\tilde P$ and since tmLQCD respects $\tilde
P$ the pion will couple to $A_1^{-+}$ and not to $A_1^{++}$.  However,
the $A_1^{++}$ channel will couple to the flavor-singlet pseudoscalar
(i.e.\ the $SU(2)$ $\eta^\prime$) and if we neglect disconnected
diagrams, then the mass of the $SU(2)$ $\eta^\prime$ is identical to
the mass of the neutral pion.  This degeneracy is not specific to the
pion-$\eta^\prime$ system; it occurs for any angular momentum irrep
with any $C$.  Therefore $\Lambda^{PC}$ contains the same spectrum of
masses for both $P$ values in neutral channels.

As discussed in Ref.~\cite{Harnett:2006fp}, all $\Lambda^{PC}$
combinations can be obtained from operators that are independent of twist angle.
Since the present work is restricted to maximal twist, those operators are not
emphasized.
In either case,
\mbox{tmLQCD} simulations will still contain mixing as discussed above:
For charged mesons we cannot separate the $\Lambda^{++}$ and $\Lambda^{--}$
pair, nor can we separate the $\Lambda^{+-}$ and $\Lambda^{-+}$ pair.
For neutral mesons we cannot separate the $\Lambda^{++}$ and $\Lambda^{-+}$
pair, nor can we separate the $\Lambda^{+-}$ and $\Lambda^{--}$ pair.
A striking example is the appearance of a pion signal for the operator
$\bar u(x)\gamma_4d(x)$.

\section{\label{sec:curvefit}Curve fitting with evolutionary algorithms}

There are several motivations for our use of evolutionary fitting.
In general, the sheer quantity of fitting to be done
requires an automated black box method, an objective being pursued by
others as well~\cite{Lin:2007iq}.  An important part of this goal is
that the method not depend on any subjective parameters such as
timestep fit ranges, a fixed number of states, or initial parameter
values.  This is not simply to speed up the fitting process but for
reproducibility as well.  While evolutionary fitting requires mutation
and breeding steps to be identified as well as probabilities to be
fixed in the algorithm, the goal of minimizing the $\chi^2/n_{dof}$ is
objective.  Any two fit algorithms, evolutionary or not, can have
their results on the same data readily compared by this statistic and
either have the fits confirmed to be equivalent or the better fit
selected.\footnote{With such comparisons, fitting algorithms
  themselves may thereby evolve.}

Other reasons for using this fitting method are specific to our
problem.  We require a method that extracts a set of common states
from multiple correlators in the same channel.  This is often
accomplished using the variational fitting
method~\cite{Michael:1985ne,Luscher:1990ck,Burch:2005wd}.  Due to the
large number of operators present, however, we run only the diagonal
correlators of the correlator matrix and not the full correlator
matrix generated by putting all possible combinations of operators in
a given channel at source and sink.  The latter would have been
required to use the variational method.  We also smear our source and
sink operators differently, which results in a correlator matrix which
is not Hermitian, which is also required for that method.
Evolutionary fitting in principle allows us to identify continuum
angular momentum states ($J$) across lattice symmetry channels which
have only octahedral symmetry.  Finally, a method that is able to
identify potentially weak contamination signals occurring due to the
use of twisted mass is also of value.

We have detailed the algorithm
itself in Refs.~\cite{von_Hippel:2007ar} and~\cite{vonHippel:2007dz}.
Here a summary is presented to provide context for evaluation of the
method.

The idea behind an evolutionary algorithm is to consider candidate
solutions to a problem as individual organisms in a population.  This
population is allowed to mutate and breed to produce successive
generations.  The types of operations involved in mutating an
individual or breeding a pair of individuals will be specific to the
problem at hand.  These must be sufficient for the organisms to have a
good probability of being able to explore the solution space.  Coupled
with a function to measure the fitness of an organism which in turn
influences which organisms will be allowed to populate the next
generation, evolution is allowed to take its course with successive
generations approaching a better solution.  Different individuals in
the population will have desirable characteristics which will be
disseminated over time with high probability to the rest of the
population.  In a sense, the population is able to explore the
solution space to the problem in parallel, making it an effective
technique in general for finding good, and sometimes unexpected,
solutions to complex problems.

In the context of lattice QCD, the organism is a fit function
that is a linear superposition of an {\em a priori} unknown number of
exponential functions whose exponents determine the spectrum of masses
in the correlator.  The simplest case involves a fit to a single
correlator and we have detailed this
elsewhere~\cite{von_Hippel:2007ar,vonHippel:2007dz}.  In the current
situation we have multiple correlators for a single channel sharing
the same energy states $\{E_m:m=1,\ldots,m_{max}\}$.  The total fit
function to all the data is then a set of fit functions, one per
correlator, each of the form:
\begin{equation}
\label{eq:multicorrfit}
G^{(i)}(t)=\sum_{n=0}^{n_{max}^{(i)}} Z_n^{(i)} \left(e^{-E_{I_n^{(i)}} t}+e^{-E_{I_n^{(i)}} (T-t)} \right)\;.
\end{equation}
Here $i$ is the correlator index, $n_{max}^{(i)}$ is the (variable)
number of states found in that correlator, $Z_n$ is the coefficient
for the energy state $E_{I_n}$, and $T$ is the temporal extent of the
lattice.
%R The following is only needed in the article.  In the thesis we could
%R add about the correlator matrix from the proceedings.
The form of the fit function reflects the fact that we are only
fitting diagonal correlators in the correlator matrix (essentially the
same operator at source and sink though we do allow for different smearings
at source and sink), and that we are using periodic
boundary conditions.  The desired best fit minimizes $\chi^2/n_{dof}$
of the total fit function $G$.
%R Put this line in thesis:
%Assuming the datasets are not correlated in any way, the new $\chi^2$
%is simply the sum of terms of the form in
%equation~(\ref{eq:chi2_single}), one per dataset.
Assuming the datasets are not correlated in any way, $\chi^2(G)$ is
simply the sum of $\chi^2(G^{(i)})$, the correlated $\chi^2$ on the
$i^{th}$ correlator~\cite{vonHippel:2007dz}.
%R The following can go in my thesis:
%The number of degrees of freedom changes from the form of
%equation~(\ref{eq:ndof_single}) to:
The number of degrees of freedom is
\begin{equation}
\label{eq:chi2_multi}
n_{dof}(G)=n_{data} - m_{max}-\sum_{i=1}^{i_{max}}n_{max}^{(i)}\;,
\end{equation}
%R Thesis:
%where $n_{data}$ now counts all included timesteps in the fit of all
%datasets.
where $n_{data}$ is the number of timesteps fit in each correlator
times the number of correlators ($i_{max}$).  The fitness of the
organism, $f(G)$, which we desire to maximize is therefore defined to
be $-\chi^2(G)/n_{dof}(G)$.  One notes that the degrees of freedom
fluctuate with the number of parameters (states and coefficients) in a
particular organism.  The complication of searching such a
discontinuous function space as well as the independence from initial
conditions (the initial population is chosen at random) are some of
the principal advantages of the evolutionary fitting method.

The information required to construct a given fit organism is coded in
its genotype.  The subfit for dataset number $i$ is represented by a
list of $n^{(i)}_{max}$ coefficients $(Z_n^{(i)},I_n^{(i)})$ where
$I\in\{1,\ldots,m_{max}\}$ is an integer index indicating to which of
the $m_{max}$ energy states $E_m$ the coefficient is associated.  In
summary, for a fit of $i_{max}$ correlators the complete genotype is
of the form:
\twocol{
\begin{eqnarray}
  \label{eq:multidatasetfit}
  \lefteqn{\mathrm{Fit\ Genotype}} \nonumber \\
  & = & ( \mathrm{Dataset\ coefficients},\mathrm{Mass\ list} ) \nonumber \\
  & = & ( (\mathrm{Dataset\ 1\ coeffs},\ldots,\mathrm{Dataset\ }i_{max}\mathrm{\ coeffs}), \nonumber \\
  &   & \mathrm{Mass\ list}  ) 
\end{eqnarray}
}
{
\begin{eqnarray}
  \label{eq:multidatasetfit}
  \lefteqn{\mathrm{Fit\ Genotype}} \nonumber \\
  & = & ( \mathrm{Dataset\ coefficients},\mathrm{Mass\ list} ) \nonumber \\
  & = & ( (\mathrm{Dataset\ 1\ coefficients},\ldots,\mathrm{Dataset\ }i_{max}\mathrm{\ coefficients}),\mathrm{Mass\ list}  ) 
\end{eqnarray}
}
with
\twocol{
\begin{eqnarray}
 \mathrm{Dataset\ }i\mathrm{\ coeffs} & = & ( (Z_1^{(i)},I_1^{(i)}), \ldots, (Z_{n_{max}^{(i)}}^{(i)},I_{n_{max}^{(i)}}^{(i)})) \nonumber \\
  \mathrm{Mass\ list} & = & (E_1, \ldots ,E_{m_{max}})\ .
\end{eqnarray}
}
{
\begin{eqnarray}
 \mathrm{Dataset\ }i\mathrm{\ coefficients} & = & ( (Z_1^{(i)},I_1^{(i)}), \ldots, (Z_{n_{max}^{(i)}}^{(i)},I_{n_{max}^{(i)}}^{(i)})) \nonumber \\
  \mathrm{Mass\ list} & = & (E_1, \ldots ,E_{m_{max}})\ .
\end{eqnarray}
}

With the above genotype defined, operations for mutation and breeding
become apparent.  The genotype is a hierarchy of lists so we coded
procedures that proceeded recursively through the structures
present.\footnote{A modern list-based language, \python~\cite{python},
  was convenient for this implementation and fast enough for our
  purposes.}  Mutations of lists involve mutating the elements in the
list.  For a list of elements of the same type, adding or removing a
random element, and if order is meaningful to the list, reordering
it, are other possible mutations.  Breeding (or crossover) of two
lists will produce two new lists containing parts of each, and
potentially of different lengths if the lists contain elements all of one
type.  Ultimately one has to mutate numbers, either reals or integers,
and this can be accomplished by adding a Gaussian random variable or
flipping bits in a binary representation respectively.  (Our integer
index has to be interpreted modulo the number of energy states
$m_{max}$ so that it always maps to an individual state.)  Breeding of
numbers can be done by randomly interpolating the real numbers or
exchanging subsets of the bits of the integers.

It is of value to introduce special mutations to the full genotype.
One mutation does a fixed number of steps of a Newtonian optimization
of the fit function of a single organism.  Here we used the
Levenberg-Marquardt~\cite{Levenberg:1944,Marquardt:1963} method with
the current functional form defined by the organism's genotype and the
values of its individual parameters as the initial conditions.  Also
we introduced a reduction mutation which would convert genotypes that
represented the same function into a common form so that the fitting
algorithm would converge to a single representation of the solution.
Consult Refs.~\cite{von_Hippel:2007ar,vonHippel:2007dz} for further detail
of these steps.

%%% Local Variables: 
%%% mode: latex
%%% TeX-master: "tmmeson"
%%% End: 

\section{\label{sec:simdetails}Simulation details}

The quenched configurations used for this work were previously
discussed in Ref.~\cite{AbdelRehim:2006ve}, and the details are
reproduced in Table~\ref{tab:params} of this work for completeness.
In addition to tmLQCD fermions, a Wilson fermion was considered for
purposes of comparison and those parameters are also listed in
Table~\ref{tab:params}.  Note that
at each of the three $\beta$ values, the four quark masses are
comparable by their approximate ratios with the physical strange quark
mass: $m_s$, $m_s/2$, $m_s/3$, and $m_s/6$.  The Wilson quark at
$\beta=6.0$ was chosen close to $m_s/2$.

The tuning to maximal twist was performed by varying $m_0$ for each $\mu_0$
until $\omega$ became $\pi/2$ as defined
by~\cite{Jansen:2003ir,AbdelRehim:2005gz,Farchioni:2004ma}
\begin{equation}\label{tuningeq}
\tan\omega=\frac{i\sum_{\vec x}\left<V_4(\vec x,t)P(0)\right>}
                 {\sum_{\vec x}\left<A_4(\vec x,t)P(0)\right>}\;,
\end{equation}
where $P,V_\mu,A_\mu$ are the local bilinears for charged mesons with
pseudoscalar, vector and axial vector quantum numbers
respectively.

\begin{table*}
\caption{\label{tab:params}The parameters used for simulations in this work.
  Lattice spacings are taken from Ref.~\protect\cite{Jansen:2003ir} using $r_0=0.5$ fm.
  Each $(am_0,a\mu_0)$ pair is the result of tuning to maximal twist
  as discussed in Ref.~\protect\cite{AbdelRehim:2006ve}, except the Wilson case
  of course.
}
\begin{ruledtabular}
\begin{tabular}{ccccclc}
$\beta$~~ & ~~$a$ [fm]~~ & \#sites & ~~\#configurations
 & ~~~~$am_0$~~~~ & ~~$a\mu_0$
& twist angle (degrees) \\
\hline
5.85 & 0.123 & $20^3\times40$ & 600 & -0.8965 & 0.0376 & 90.0$\pm$0.3 \\
     &       &                &     & -0.9071 & 0.0188 & 90.2$\pm$0.6 \\
     &       &                &     & -0.9110 & 0.01252 & 90.6$\pm$0.8 \\
     &       &                &     & -0.9150 & 0.00627 & 90.6$\pm$1.6 \\
\hline
6.0  & 0.093 & $20^3\times48$ & 600 & -0.8110 & 0.030 & 90.4$\pm$0.4 \\
     &       &                &     & -0.8170 & 0.015 & 91.0$\pm$0.7 \\
     &       &                &     & -0.8195 & 0.010 & 92.5$\pm$1.0 \\
     &       &                &     & -0.8210 & 0.005 & 95.5$\pm$2.1 \\
\cline{5-7}
     &       &                &     & -0.7835 & 0.0   & (Wilson) \\
\hline
6.2  & 0.068 & $28^3\times56$ & 200 & -0.7337 & 0.021649 & 89.1$\pm$0.8 \\
     &       &                &     & -0.7367 & 0.010825 & 87.3$\pm$1.8 \\
     &       &                &     & -0.7378 & 0.007216 & 86.3$\pm$2.8 \\
     &       &                &     & -0.7389 & 0.003608 & 86.4$\pm$4.5
\end{tabular}
\end{ruledtabular}
\end{table*}

The diagonal correlators of the $384$ extended and $16$ local
operators from Section~\ref{sec:latticeops} were calculated for each
lattice spacing using degenerate quarks at our four quark masses in
both the charged and neutral channels for twisted mass and at the
single quark mass for Wilson.  Neutral twisted mass channels did not
include the so-called disconnected contributions,
i.e.\ contractions between a quark
and anti-quark within the source (or sink) operator.

Smearing was used to reduce contributions from excited states 
to correlators at the shortest time extension.  
Gaussian quark smearing was
performed at the sink and stout link smearing at both source and sink
as described in Ref.~\cite{AbdelRehim:2006qu}.  Quark smearing parameters
were $\alpha=0.15$ and $n_{\alpha}=64$ for all lattice spacings.
Stout link smearing used $\rho=0.15$ for $\beta=5.85$ and $\rho=0.2$ for
$\beta=6.0$ and $6.2$ and $n_\rho=16$ for all three lattice
spacings.\footnote{See Ref.~\cite{AbdelRehim:2006qu} for explicit
  definitions of these parameters.}
For comparison, correlators of unsmeared operators were also computed.

Before fitting, diagonal correlators corresponding to operators that
differed only in their row $\irrepi$ were averaged since these must be
the same statistically by symmetry.  Specifically, the diagonal
correlators from operators in Equation (\ref{eq:finalsymops}) with the
same $(\irrep^{PC},\irrepdirac^{\irrepcopy_\diraclower P_\diraclower
  C_\diraclower} , \irreplink^{P_\linklower C_\linklower})$ were
averaged, leaving the number of correlators to be fit for each channel
given by Table~\ref{tab:totalreduction} for a fixed quark mass, fixed type
(twisted mass charged, twisted mass neutral, and Wilson), fixed
lattice spacing, and fixed smearing.

For each such set of correlators, evolutionary fits were done to the
data at all timesteps using an overall population of $480$
organisms.\footnote{The population was distributed over $4$ islands
  with parameters $N_{elite}=5$, $N_{diversity}=5$, $N_{mutant}=20$.
  See Refs.~\cite{von_Hippel:2007ar,vonHippel:2007dz}.}  Two fits were done
for each dataset to test fit consistency. The first was stopped at exactly
$600$ generations.  The second run, the results of which were
used, performed at least $600$ generations but was allowed to continue
up to $1200$ generations, stopping in between only if no improvement
in the best genotype of a given generation was seen for $200$
generations.  The genotype was allowed to contain up to $8$ masses,
and the fit for each correlator could have up to $8$ coefficients
pointing to elements of the mass list.  All coefficients and masses
were restricted to be positive.  Once the evolutionary algorithm had
obtained a best fit function for the dataset, bootstrap
errors~\cite{efron82} were generated by fitting this function to
bootstrap configurations.\footnote{Bootstrap fits were done using the
  Levenberg-Marquardt method~\cite{Levenberg:1944,Marquardt:1963} with
  the fit functional form and the initial parameters taken from the
  best fit found by the evolutionary algorithm on the actual data.}

\section{\label{sec:results}Results}

\subsection{\label{subsec:obscurvefit}Observations on curve fitting}

\begin{figure}
\includegraphics[width=.45\textwidth,keepaspectratio=,clip=]{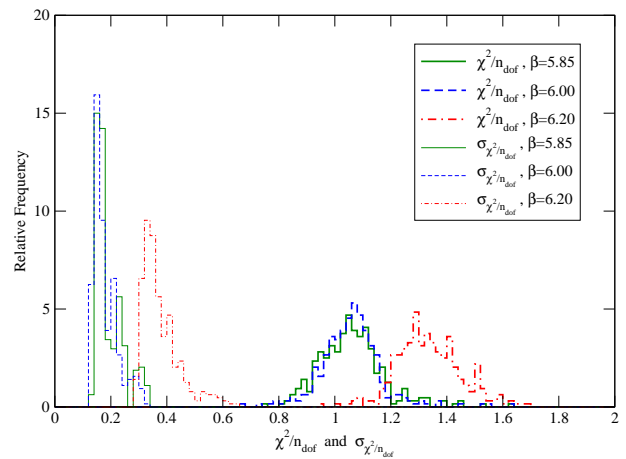}
\caption{\label{fig:chi2perdof}Shown are normalized histograms of the
  $\chi^2/n_{dof}$ (thick lines) of all the channels fit for each of
  our lattice spacings.  To the left of the diagram one also has
  histograms of the standard deviation of these values,
  $\sigma_{\chi^2/n_{dof}}$ (thin lines).  Smeared and unsmeared data
  at the same lattice spacing had essentially the same histograms so
  the plot does not distinguish them.}
\end{figure}

To assess the ability of the fitting algorithm to obtain accurately
good fits to our data, we present in Figure~\ref{fig:chi2perdof} a
histogram of the $\chi^2/n_{dof}$ for all $320$ twisted mass fits at
each lattice spacing.\footnote{These fits were for each of the $20$
  $\Lambda^{PC}$ channels at four degenerate quark masses with
  correlators both neutral and charged, smeared and unsmeared.  Later
  we also did fits to the combined $E$ and $T_2$ channels as well as
  for the $A_2$ channel combining smeared and unsmeared correlators.
  There, fit quality statistics were in line with these, with
  $\beta=6.2$ once again systematically higher.}  While the plot
clearly shows that all the fits are reasonable, falling largely in the
range $0.9-1.4$, the fits for the finest lattice spacing, $\beta=6.2$,
while having the same shape of distribution, have their mean shifted
upward from that of the coarser spacings by about $0.3$.  At the finer
lattice spacing more states are resolved so one possibility would be
that the fitting algorithm has not had sufficient time to find the
true minima.  However, in comparing these fits that were allowed to go
up to $1200$ generations with those required to stop at $600$
generations, no marked improvement in the histogram is seen.  A good
fit for all our data was achieved at $600$ generations so
this is not the source of the discrepancy.

The actual source of this discrepancy can be traced to lower
statistics, as our $\beta=6.2$ data involved only $200$ configurations
compared to $600$ at the coarser lattice spacings.  In
Ref.~\cite{von_Hippel:2007ar} we fit subsets of the configurations of our
$\beta=6.0$ data to see what the effect of poor statistics would be
on the result.  The subfit on only $200$ configurations shown in that
paper clearly demonstrates an increase in the $\chi^2/n_{dof}$
comparable to what we find here with our $\beta=6.2$ data.  Since our
poorer $\beta=6.2$ fits can be explained by fewer statistics, we
conclude that where good fits to the data exist they are found by the
fitting algorithm.  It is notable that the shape of this histogram can
be used to assess the minimum number of configurations required for a
proper simulation.  Presumably, once a sufficient number of
configurations have been used that states are reliably identified, one
will get a good distribution of $\chi^2/n_{dof}$ centered on
one.\footnote{Obviously the number of correlators fit in each channel
  varied greatly so the width of the distribution is partly related to
  the statistics induced by this variation.}  A greater number of
configurations would then serve to lower the error on the states found
and narrow the overall distribution.\footnote{One may wonder why our
  smeared and unsmeared correlators had comparable histograms if the
  smeared data had to resolve a fewer number of states.  Why this
  should be the case is likely due to the asymmetrical smearing that
  is done to our source and sink operators.  While it is true that we
  are removing excited states by smearing, in our case we are
  producing noisier correlators with the result that our statistics
  remain wanting for our smeared correlators at our finest lattice
  spacing.  Our baryon study with the same smearing showed a similar
  result~\cite{AbdelRehim:2006qu}.}

Since the errors of the parameters were obtained by bootstrapping the
functional form found by the evolutionary algorithm, it was also
possible to calculate the variation in the $\chi^2/n_{dof}$ for each
fit.  Figure~\ref{fig:chi2perdof} includes histograms of the standard
deviation $\sigma_{\chi^2/n_{dof}}$ of our fits.  For the coarser
lattice spacings with $600$ configurations one finds that there is a
relatively narrow distribution of the $\chi^2/n_{dof}$ across
bootstrap configurations.  This gives some confidence that the fitting
algorithm has found a stable functional form for the fit.  For
$\beta=6.2$, however, the variation is found to be wider.  That this
is due to the fewer configurations available at this spacing can also
be confirmed by consulting Ref.~\cite{von_Hippel:2007ar} which shows that
our subfit to $200$ configurations displayed a similar trend.
Similar findings are obtained by plotting the standard quality of fit $Q$,
as discussed in Ref.~\cite{Robsthesis}.

\begin{figure}
\includegraphics[width=.45\textwidth,keepaspectratio=,clip=]{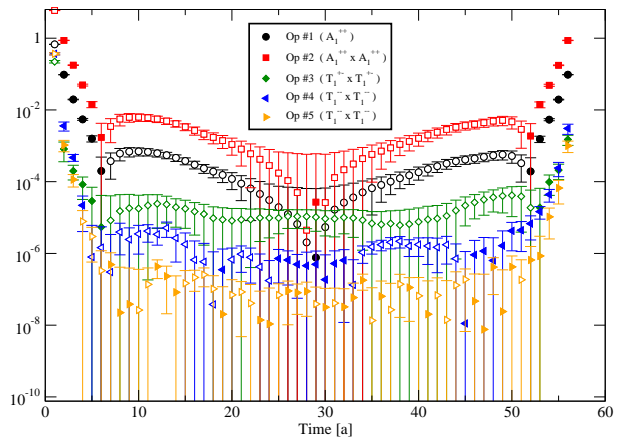}
\caption{\label{fig:ppA1corrs} Shown are the five unsmeared diagonal
  correlators for the charged $A_1^{++}$ channel at $\beta=6.2$ and
  lightest quark mass.  Solid points are positive values and open are
  negative.
  In parentheses after each number the octahedral irrep of the local
  quark structure and the gauge field used to generate the operator
  are given.  The first operator has no extended gauge field
  structure.  The last two operators use different $T_1^{--}$ local
  quark representations in their construction, operator $\#4$ using
  quark structure $T_1^{2--}$, with the superscript $2$ indicating the
  multiplicity $\alpha$ from Table~\ref{tab:spinorreduction}.}
\end{figure}

Because our evolutionary algorithm fitting function was designed to
fit correlators that are sums of decaying exponentials only, this
causes a systematic error in the $A_1^{++}$ channel where, due to a
quenching artifact, this is not the functional form of the correlator.
In quenched lattice QCD one has a ghost contribution in the scalar
correlator due to the $\eta'$-$\pi$ intermediate state being light and
of negative norm in this approximation~\cite{Bardeen:2001jm}. In our
data this effect is most pronounced at the lightest quark mass in the
$\beta=6.2$ charged channel.  See Figure~\ref{fig:ppA1corrs} where
our five correlators in this channel are plotted. Because we only have
good $C$ and product $PG$ in our neutral and charged channels
respectively, this artifact will appear in other $A_1$
channels~\cite{AbdelRehim:2006ve}. Because the coefficients in our fit
functions are constrained to be positive, the effect of negative ghost
contributions is primarily to produce a poor fit.  Several of the
higher outliers of $\chi^2/n_{dof}$ in Figure~\ref{fig:chi2perdof} are
readily traced to $A_1$ channels containing this ghost contribution.

In order to evaluate the consistency of the fitting algorithm we did
at least two runs in each channel and found no significant variation
in the results.  Where difference occurred it would tend to be a
single state with large error found in a single correlator.  Given a
finite number of generations in a run this is to be expected.

The fitting algorithm appears to have resolved states predictably.
Unsmeared channels found greater numbers of states than did smeared channels.
In some
channels no state was found, which affirms that the method does
not claim signals in everything.  Occasionally a clearly spurious state
with large error would appear in a single correlator which is also
expected from the statistical variation of the data itself.  Finer
lattice spacings in the unsmeared case found more states since a
greater number of excitations could be resolved.  Occasionally the
$\beta=6.2$ lattice spacing appeared to resolve an intermediate state
not seen in the other spacings.  While the bifurcation of states
%L should be expected at finer spacing, the fewer configurations in this
%L channel make it unclear whether this was occurring.
could certainly occur at finer spacing, the fewer configurations at
$\beta=6.2$ do not allow us to make a strong conclusion from this observation.

Occasionally states of low energy would appear in some channels with
large error below the obvious ground state.  Such errant signals are a
consequence of fitting at Euclidean times far from the source, where
correlators are noisy.  They may be an artifact of time step
correlation not being entirely removed through the use of the
correlated $\chi^2$.  We regard the presence of these noisy spurious states
as a reminder that we are using a true black box method: the $\chi^2/n_{dof}$
is slightly reduced by their presence, and we have not prevented their
appearance through human intervention such as choosing a fitting window in
Euclidean time.

As found in Ref.~\cite{von_Hippel:2007ar}, the computer time required for an
evolutionary fit scales with the number of parameters required.  This
depends not only on the number of states in the channel but includes
the coefficients required on each correlator, and hence depends on the
number of correlators fit in the channel.  By smearing we reduce the
number of states and therefore the number of coefficients and thus
smearing allowed us to fit many correlators in the same channel
quickly.

As part of the analysis we tried to fit all
smeared correlators corresponding to the $J=4$ channel which includes
all four octahedral irreps ($A_1$, $E$, $T_1$, and $T_2$) to which its
subduced representation reduces for a given $PC$.  Such fits involved fitting on the order of forty
correlators simultaneously and, although taking proportionately
longer, were, in the end, tractable by the fitting algorithm.

%%% Local Variables: 
%%% mode: latex
%%% TeX-master: "tmmeson"
%%% End: 

\subsection{\label{subsec:obstwistedmass}Observations on twisted mass}

\begin{figure*}
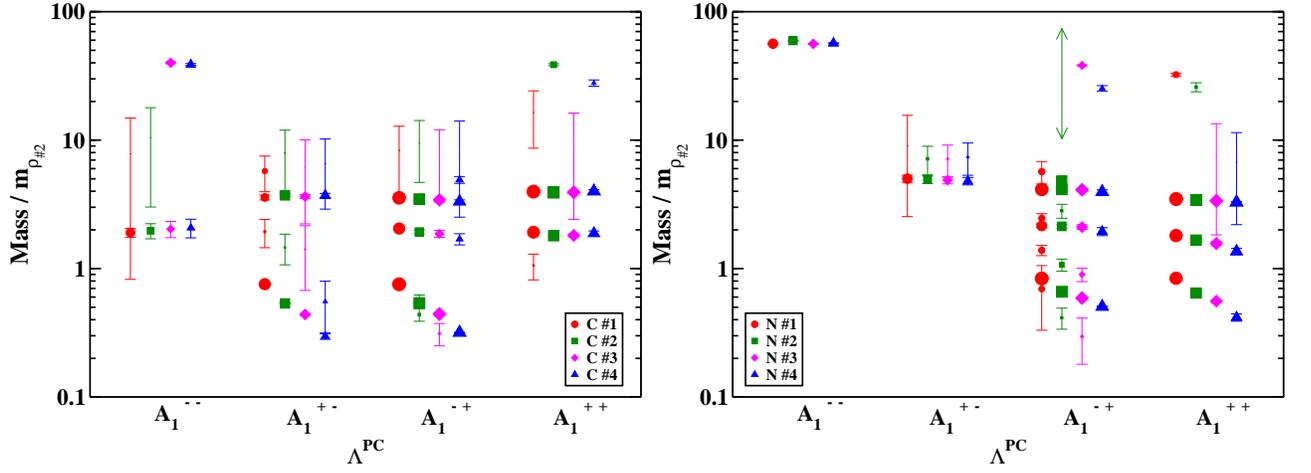

\includegraphics[width=.47\textwidth,keepaspectratio=,clip=]{multimassA1C.eps}
\includegraphics[width=.47\textwidth,keepaspectratio=,clip=]{multimassA1N.eps}
\caption{\label{fig:multimassA1}The results of the fits to
  the four $A_1$ channels at $\beta=5.85$ for the unsmeared correlators.
  For each of the charged~(C) and neutral~(N) channels,
  there are four fits for the quark masses going from the heaviest
  (mass~$\#1$) to the lightest (mass~$\#4$).  Points are scaled to the
  $\rho$ mass found in the smeared charged mass $\#2$ channel. Point
  sizes are scaled to reflect their significance as given by
  Equation (\ref{eq:scaling}).}
\end{figure*}

\begin{figure*}
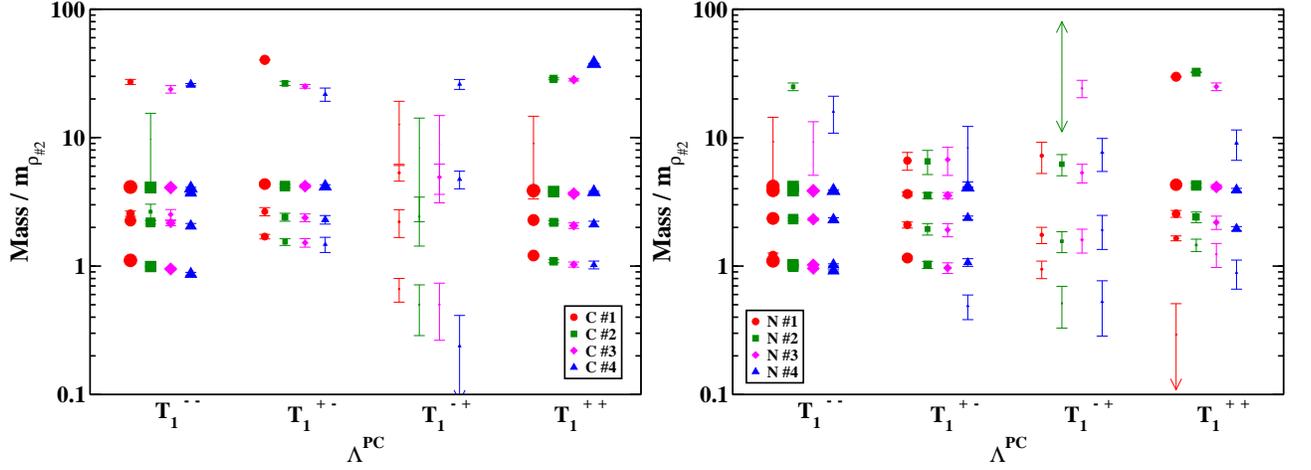

\includegraphics[width=.47\textwidth,keepaspectratio=,clip=]{multimassT1C.eps}
\includegraphics[width=.47\textwidth,keepaspectratio=,clip=]{multimassT1N.eps}
\caption{\label{fig:multimassT1}The results of the fits to
  the four $T_1$ channels at $\beta=5.85$ for the
  unsmeared correlators.  Notation is identical with
  Figure~\ref{fig:multimassA1}.}
\end{figure*}

Figures~\ref{fig:multimassA1} and \ref{fig:multimassT1} show fits with
both charged and neutral operators in the $A_1$ and $T_1$ channels,
which couple to spin zero and one respectively in the continuum.  Results
at four different degenerate quark masses are shown.
The sizes of the points are scaled to give a qualitative impression of
the significance of the state in the datasets via the factor:
\begin{equation}
  \frac{2}{\pi}\arctan\left(\frac{1}{N}\sum_{i=1}^N \frac{Z^{(i)}}{\epsilon^{(i)}}\right)\;.
  \label{eq:scaling}
\end{equation}
Here $N$ is the number of correlators involved in the fit, $Z^{(i)}$ is
the coefficient on the $i^{th}$ correlator for the given energy state
(taken to be zero if there was no coefficient found on that dataset)
and $\epsilon^{(i)}$ is the bootstrap error of the given coefficient.  The
$\arctan$ function is used to ensure the scale factor ranges from $0$
to $1$.  The utility of the scaling is that it facilitates the
identification of the same state across lattice spacings and quark
masses for extrapolation purposes.  The scale factor is not
numerically involved in any extrapolation however.

In Figures~\ref{fig:multimassA1} and \ref{fig:multimassT1}, one is
able to see easily the effect of the twisted mass channel
contamination discussed in Section~\ref{sec:twistedmass}.  In
Figure~\ref{fig:multimassA1} the ground state pseudoscalar (``pion'')
is clearly visible across quark masses where it should be in the
$A_1^{-+}$ channel for both the charged and neutral correlators.
For the charged operators,
channels are expected to contaminate others with the same $PG$
product.  The charged pion shows up clearly as contamination in the
$A_1^{+-}$ channel as predicted.  That it is contamination is clear
not only because there is no corresponding signal in the neutral
channel but also the larger error and smaller symbols are both
indicative of a weaker signal than the authentic one in the $A_1^{-+}$
channel.  Comparison of the $A_1^{++}$ channel with the $A_1^{--}$
channel similarly shows contamination of the authentic scalar
$A_1^{++}$ ground state signal in its twinned channel.

Turning to the neutral case, charge conjugation is respected and one
sees the pion now contaminating the $A_1^{++}$ channel.  States in the
latter channel in turn are contaminating the $A_1^{-+}$ channel where
one clearly sees weak contamination states interspersed with authentic
states.  It is a testament to the fitting algorithm that it is able to
distinguish these errant states.

The strength of the ground state vector (``$\rho$ meson'') allows us
to confirm similarly the contamination relations for those channels
not evident in the scalar case.  The $\rho$ is clearly identified in
both charged and neutral $T_1^{--}$ channels in
Figure~\ref{fig:multimassT1}.  Consideration of the contamination in
the charged channel suggests the particle should appear in the
$T_1^{++}$ channel, which it clearly does.  In the neutral case the
$\rho$ contamination is appearing in the $T_1^{+-}$ channel as
expected.  It is worth noting that contamination states are not
appearing in unexpected channels.  The strength of the pion and the
$\rho$ as signals, even in contamination, would make them easily
discernible.

When contamination is weak or when the actual states in the channel
being contaminated are themselves weak one may see distortions in the
actual state if the two states are nearby or cannot be distinguished.
Higher statistics are required to disentangle correctly the greater
number of states that will appear in a given channel. The value of
being able to compare neutral and charged channels which get
contaminated differently is readily apparent.

Overall, contamination appears to decrease with finer lattice spacing.
At lighter quark mass the contamination also lessens but so does the
actual signal in the original channel.  In our simulation, smearing
reduced the contamination strength and in our smeared fits it was only
possible to identify reliably ground state $\rho$ and pion contamination.
The decrease in contamination with lattice
spacing and the effect of smearing can be seen in the fit plots in
Section~\ref{sec:spectroscopy}.
The fits in Figures~\ref{fig:multimassA1} and
\ref{fig:multimassT1}, being to data both unsmeared and at our coarsest
lattice spacing, are for illustration of the contamination effect.  The
smeared data we actually fit for our results appear to have limited
contamination identified by the fitting algorithm.

%%% Local Variables: 
%%% mode: latex
%%% TeX-master: "tmmeson"
%%% End: 

\subsection{\label{sec:spectroscopy}Meson spectroscopy}

\begin{figure*}[p]
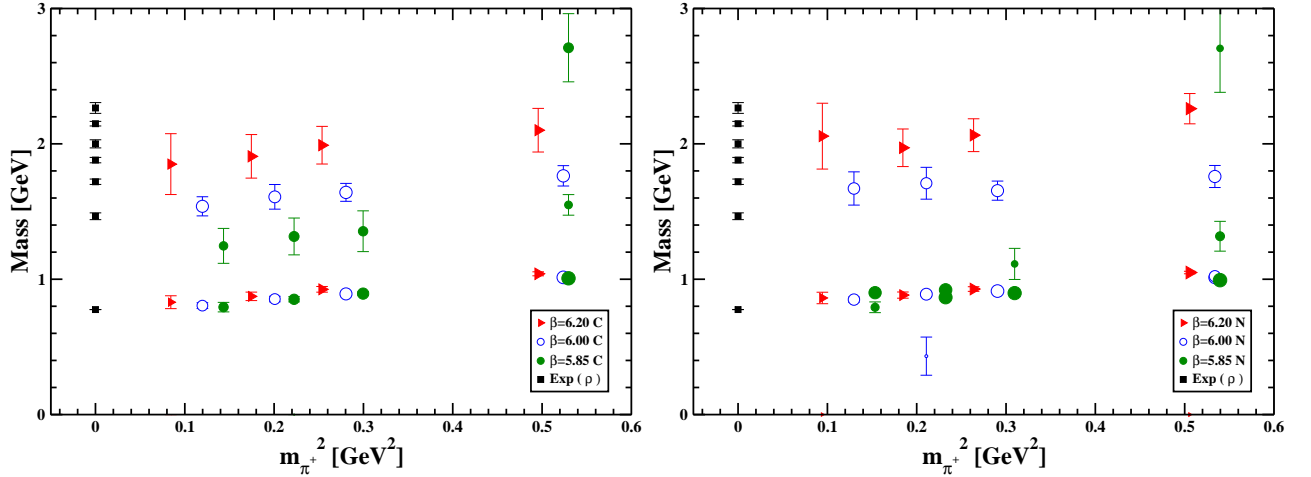

\includegraphics[width=.47\textwidth,keepaspectratio=,clip=]{t1mmC.eps}
\includegraphics[width=.47\textwidth,keepaspectratio=,clip=]{t1mmN.eps}
\caption{\twocol{}{\scriptsize}\label{fig:mmT1}Fits to smeared
  $T_1^{--}$ correlators are shown.  Experimental measurements~\cite{Yao:2006px}
  are plotted for comparison.}
\end{figure*}

\begin{figure*}[p]
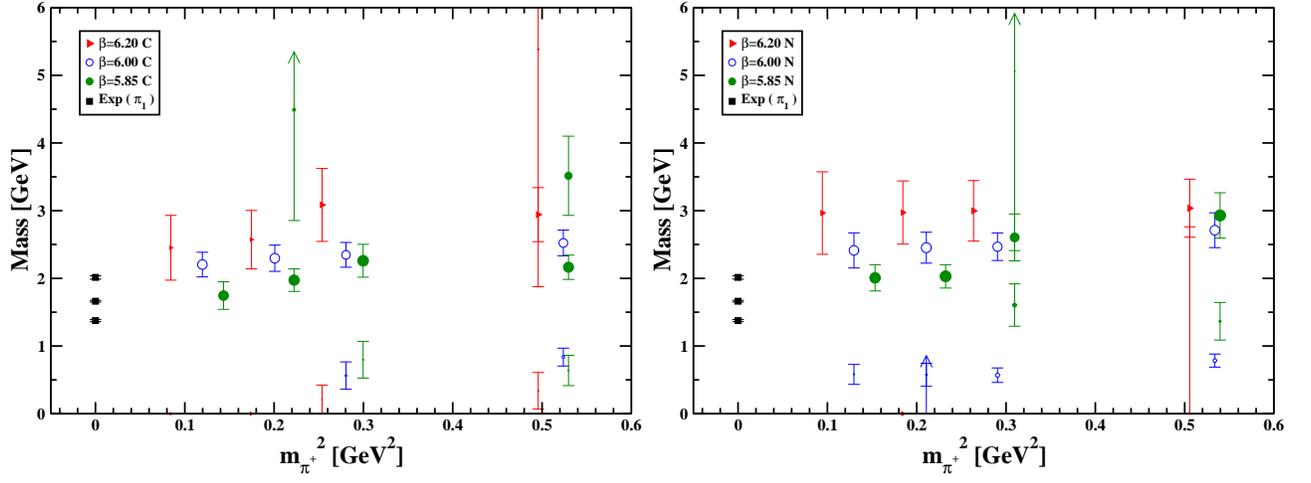

\includegraphics[width=.47\textwidth,keepaspectratio=,clip=]{t1mpC.eps}
\includegraphics[width=.47\textwidth,keepaspectratio=,clip=]{t1mpN.eps}
\caption{\twocol{}{\scriptsize}\label{fig:mpT1}Fits to smeared
  $T_1^{-+}$ correlators are shown.  A clear signal appears between
  $2$ and $3$ GeV which has been extrapolated.  The lower points
  seem to suggest that the fitting algorithm is detecting some other
  state, likely channel contamination due to tmLQCD.  This is an
  exotic channel.  Experimental claims~\cite{Yao:2006px} are plotted for
  comparison.}
\end{figure*}

The data from the smeared $T_1^{--}$ and $T_1^{-+}$ channels are provided in
Figures~\ref{fig:mmT1} and \ref{fig:mpT1} respectively.
The sizes of all data points are scaled by Eq.~(\ref{eq:scaling}) so tiny
points are often devoid of any physics content.
Similar plots for all other $\Lambda^{PC}$ appear in Ref.~\cite{Robsthesis}.

In Figure~\ref{fig:mmT1}, the extrapolated ground state
  vector meson (the $\rho$) is in agreement with its physical value, and
  a single excited state is resolved.  The neutral $\beta=5.85$
  excited state, while present at every quark mass, appears to be
  interfering with the ground state.  The resulting poor chiral
  extrapolations of the neutral $\beta=5.85$ data will be
  excluded from the continuum extrapolation of both the ground and
  excited states.  At the heaviest quark mass the ground state for the
  neutral $\beta=6.0$ fit is bifurcated (the points are
  indistinguishable on this plot) presumably because
  statistical variation made this a slightly more probable fit.
Due to limited statistics in the $\beta=6.2$ channel (recall
Section~\ref{subsec:obscurvefit}), there may be increased systematic errors
that arise during the fitting at $\beta=6.2$.  One might therefore prefer to
omit $\beta=6.2$ data from continuum extrapolations.

In Figure~\ref{fig:mpT1}, results in the exotic $T_1^{-+}$ channel are
plotted.  A clear signal is evident between 2 and 3 GeV, while tiny points
also appear in the small mass range.  These tiny points may be attributed
in part to contamination with the other $T_1$ channels due to tmLQCD's
symmetry structure, and in
part to statistical variation within the fitting procedure.
At our largest quark mass, which is approximately the strange
quark mass, our results are consistent with the lattice study
of Ref.~\cite{Lacock:1996vy} which used displaced quarks to generate angular
momenta in the operators for quenched simulations at $\beta=6.0$.

\begin{figure}
\includegraphics[width=.45\textwidth,keepaspectratio=]{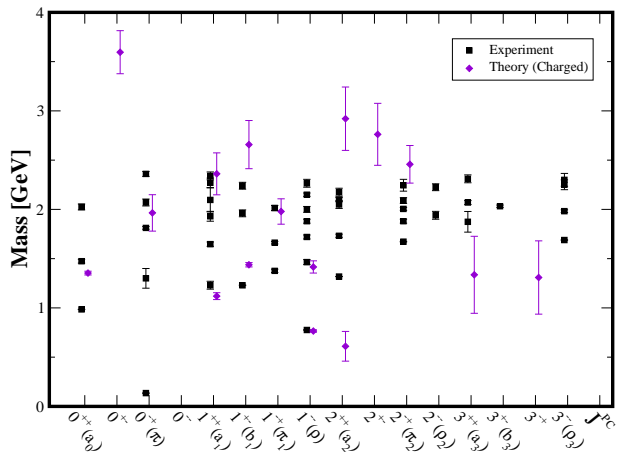}
\caption{\label{fig:final} Experimental values~\cite{Yao:2006px}
  of light unflavored
  mesons (squares) in the isovector channel for each $J^{PC}$ are
  compared to the results of this simulation (diamonds) that are the
  weighted average over lattice spacings $\beta=5.85$ and $\beta=6.0$
  extrapolated to zero quark mass.}
\end{figure}

The results of all our simulations, after linear chiral extrapolations,
are given
in Table~\ref{tab:extrapolations} along with continuum extrapolations.
Extrapolations that omit the $\beta=6.2$ data, i.e.\
simple averages of data at the other two $\beta$ values, are provided in
Table~\ref{tab:coarseaverage} for all $\Lambda^{PC}$.
The predictions for charged meson masses are also shown graphically in
Figure~\ref{fig:final}, together with the experimental mass spectrum.

\twocol{}{\begingroup\singlespacing}
\begin{table*}
\begin{ruledtabular}
\caption{\label{tab:extrapolations} Extrapolations of each resolved
  channel labeled by its irrep ($\Lambda^{PC}$) and its type, charged~(C)
  or neutral~(N), are given.  States designated by an asterisk
  indicate a higher energy state in the same channel.  The inferred
  continuum quantum number $J$ of the state is also given. Chiral
  extrapolations (linear) for each $\beta$ are shown along with their
  $\chi^2/n_{dof}$ in subsequent columns.  The continuum extrapolation
  of these results is shown in the last column.  We
  have omitted the pion ($A_1^{-+}$) from the table.  All energies are in GeV.}
\newcolumntype{d}{D{.}{.}{6}}
\begin{tabular}{lccdddddddd}
\multicolumn{3}{c}{Channel} & \multicolumn{2}{c}{$\beta=5.85$} & \multicolumn{2}{c}{$\beta=6.0$} & \multicolumn{2}{c}{$\beta=6.2$} & \multicolumn{2}{c}{\mbox{Final}} \\
\multicolumn{1}{c}{Irrep(s)}& \multicolumn{1}{c}{J}&\multicolumn{1}{c}{C/N} & \multicolumn{1}{c}{Energy} & \multicolumn{1}{l}{$\chi^2/\nu$} & \multicolumn{1}{c}{Energy} & \multicolumn{1}{l}{$\chi^2/\nu$} & \multicolumn{1}{c}{Energy} &  \multicolumn{1}{l}{$\chi^2/\nu$} & \multicolumn{1}{c}{Energy} & \multicolumn{1}{l}{$\chi^2/\nu$}\\ \hline
$A_1^{++}$ & 0 & C & 1.35(3) & 1.579 & 1.36(3) & 0.131 & 1.40(6) & 0.015 & 1.40(6) & 0.231 \\
          & & N & 1.01(4) & 2.566 & 1.21(4) & 0.198 & 1.20(8) & 0.012 & 1.39(8) & 1.683 \\
$A_1^{+-}$ & 0 & C & 3.3(3) & 0.015 & 4.2(4) & 1.576 & 5.1(10) & 0.023 & 5.6(8) & 0.053 \\
          & & N & 3.5(4) & 0.688 & 2.5(3) & 0.014 & 3.1(6) & 0.194 & 2.1(7) & 2.209 \\
$A_1^{*-+}$ & 0 & C & 0.31(17) & 2.815 & 1.96(18) & 0.048 & 2.08(10) & 0.303 & 2.2(3)\footnotemark[1] & ... \\
          & & N & 1.47(16) & 0.027 & 1.69(10) & 0.268 & 1.96(13) & 1.548 & 2.14(19) & 0.309 \\
$T_1^{++}$ & 1 & C & 0.98(5) & 0.371 & 1.25(5) & 0.990 & 1.6(2) & 0.376 & 1.65(12) & 0.461 \\
          & & N & 1.33(2) & 0.101 & 1.428(17) & 0.003 & 1.52(10) & 0.010 & 1.57(5) & 0.061 \\
$T_1^{*++}$ & 1 & C & 2.4(3) & 0.843 & 2.3(3) & 0.213 & 3.3(6) & 0.315 & 2.9(6) & 1.688 \\
          & & N & 3.0(7) & 0.109 & 2.4(3) & 0.298 & 5.8(13) & 0.459 & 2.9(10) & 7.283 \\
$T_1^{+-}$ & 1 & C & 1.40(3) & 1.507 & 1.48(3) & 1.018 & 1.74(8) & 0.961 & 1.69(7) & 4.136 \\
          & & N & 1.14(6) & 0.715 & 1.62(6) & 0.018 & 1.57(7) & 0.199 & 1.90(10) & 6.728 \\
$T_1^{*+-}$ & 1 & C & 2.6(4) & 1.095 & 2.7(3) & 0.535 & 2.8(6) & 0.011 & 2.9(7) & 0.007 \\
          & & N & 3.1(4) & 0.283 & 2.8(4) & 0.021 & 3.2(4) & 1.833 & 3.1(6) & 0.591 \\
$T_1^{-+}$ & 1 & C & 1.80(18) & 0.774 & 2.15(18) & 0.008 & 2.4(4) & 0.258 & 2.7(4) & 0.022 \\
          & & N & 1.7(2) & 0.566 & 2.3(2) & 0.027 & 2.9(5) & 0.000 & 3.3(5) & 0.157 \\
$T_1^{--}$ & 1 & C & 0.767(19) & 0.141 & 0.764(16) & 0.036 & 0.79(3) & 0.066 & 0.78(3) & 0.573 \\
          & & N & 0.797(6) & 0.508 & 0.812(12) & 0.515 & 0.80(2) & 0.122 & 0.79(5)\footnotemark[1] & ... \\
$T_1^{*--}$ & 1 & C & 1.18(12) & 0.006 & 1.50(7) & 0.030 & 1.82(18) & 0.021 & 2.00(19) & 0.253 \\
          & & N & 0.84(2) & 3.926 & 1.63(10) & 0.199 & 1.88(16) & 0.154 & 2.2(4)\footnotemark[1] & ... \\
$(E,T_2)^{++}$ & 2 & C & 0.4(2) & 0.340 & 0.75(19) & 0.290 & 1.0(4) & 0.244 & 1.3(4) & 0.003 \\
          & & N & 0.6(4) & 0.049 & 0.52(13) & 0.179 & 0.5(3) & 0.051 & 0.4(4) & 0.002 \\
$(E,T_2)^{*++}$ & 2 & C & 3.7(7) & 0.196 & 2.7(4) & 0.063 & 2.9(9) & 0.127 & 1.9(10) & 0.468 \\
          & & N & 2.4(3) & 2.235 & 3.0(4) & 0.103 & 3.2(8) & 0.037 & 3.7(8) & 0.023 \\
$(E,T_2)^{+-}$ & 2 & C & 2.8(6) & 0.057 & 2.7(4) & 0.065 & 2.9(8) & 0.018 & 2.8(9) & 0.066 \\
          & & N & 3.2(5) & 0.005 & 2.9(4) & 0.031 & 3.1(6) & 0.026 & 2.9(8) & 0.172 \\
$(E,T_2)^{-+}$ & 2 & C & 2.3(2) & 0.334 & 2.9(4) & 0.025 & 3.3(9) & 0.121 & 3.7(7) & 0.001 \\
          & & N & 2.4(2) & 4.133 & 2.8(4) & 0.001 & 3.0(7) & 0.344 & 3.2(7) & 0.001 \\
$A_2^{++}$ & 3 & C & 0.5(7) & ... & 1.8(5) & 0.002 & 3.5(8) & 0.052 & 4.3(10) & 0.573 \\
          & & N & 0.7(4) & ... & 10.(8) & 0.005 & 3.2(8) & 0.017 & 4.4(11) & 0.912 \\
$A_2^{-+}$ & 3 & C & 1.3(4) & 0.230 & 1.4(7) & 0.672 & 1.1(6) & 0.072 & 1.1(8) & 0.113 \\
          & & N & 1.5(5) & 0.376 & 1.2(4) & 0.088 & 1.2(4) & 0.664 & 1.0(6) & 0.009 \\
\end{tabular}
\footnotetext[1]{Continuum extrapolation excluded $\beta=5.85$ data point.}

\end{ruledtabular}
\end{table*}
\twocol{}{\endgroup}

\twocol{}{\begingroup\singlespacing}
\begin{table}
\caption{\label{tab:coarseaverage} Weighted average of the
  chirally extrapolated $\beta=5.85$ and $\beta=6.0$ lattice spacing
  results of each resolved channel labeled by its irrep
  ($\Lambda^{PC}$) and its type (charged or neutral) are given.
  States designated by an asterisk indicate a higher energy state in
  the same channel.  The inferred continuum quantum number $J$ of the
  state is also given. We have omitted the pion ($A_1^{-+}$) from the table.
  All energies are in GeV.}
\begin{ruledtabular}
\begin{tabular}{lcdd}
\multicolumn{2}{c}{Channel} & \multicolumn{2}{c}{\mbox{Energy}} \\
\multicolumn{1}{c}{Irrep(s)}& \multicolumn{1}{c}{J}& \multicolumn{1}{c}{~~~~~~~Charged} & \multicolumn{1}{c}{~~~~~~~Neutral} \\ \hline
$A_1^{++}$ & 0 & 1.353(19)
& 1.12(3) \\
$A_1^{+-}$ & 0 & 3.6(2)
& 2.9(3) \\
$A_1^{*-+}$ & 0 & 1.96(18)\footnotemark[1]
& 1.63(8) \\
$T_1^{++}$ & 1 & 1.12(3)
& 1.394(14) \\
$T_1^{*++}$ & 1 & 2.4(2)
& 2.5(2) \\
$T_1^{+-}$ & 1 & 1.44(2)
& 1.37(4) \\
$T_1^{*+-}$ & 1 & 2.7(2)
& 2.9(3) \\
$T_1^{-+}$ & 1 & 1.98(13)
& 1.98(16) \\
$T_1^{--}$ & 1 & 0.765(12)
& 0.812(12)\footnotemark[1] \\
$T_1^{*--}$ & 1 & 1.42(6)
& 1.63(10)\footnotemark[1] \\
$(E,T_2)^{++}$ & 2 & 0.61(15)
& 0.53(13) \\
$(E,T_2)^{*++}$ & 2 & 2.9(3)
& 2.6(3) \\
$(E,T_2)^{+-}$ & 2 & 2.8(3)
& 3.0(3) \\
$(E,T_2)^{-+}$ & 2 & 2.46(19)
& 2.52(20) \\
$A_2^{++}$ & 3 & 1.3(4)
& 0.7(4) \\
$A_2^{-+}$ & 3 & 1.3(4)
& 1.3(3) \\
\end{tabular}
\footnotetext[1]{Excluded $\beta=5.85$ data point.}

\end{ruledtabular}
\end{table}
\twocol{}{\endgroup}

Note that our results from the heaviest quark mass at $\beta=6.0$
(provided explicitly in Ref.~\cite{Robsthesis})
can be compared directly to Ref.~\cite{Lacock:1996vy}, and good agreement is
found for the available channels: $a_0$, $\pi$, $\rho$, $b_1$, $a_1$ and $a_2$.
The exotic states which we identify in the $0^{+-}$, $1^{-+}$, and $2^{+-}$
channels, for the strange quark mass at $\beta=6.0$, are also found to be
comparable to values which they ascribe to those same $J^{PC}$.

Our result for the $1^{-+}$ from Table~\ref{tab:coarseaverage}
agrees well with the general consensus of
quenched~\cite{Lacock:1996vy,Bernard:1997ib,Mei:2002ip,Hedditch:2005zf,Cook:2006hh}
and dynamical~\cite{Lacock:1998be,Bernard:2003jd}
lattice studies that the lightest exotic meson is $1^{-+}$ with a mass
near 1.9 GeV.
We also note that the flux-tube model predicts
exotic hybrids $0^{+-}$, $1^{-+}$, and $2^{+-}$ near $1.9$ GeV in the
light isovector channel~\cite{Isgur:1985vy}.

Although results of our simulations with Wilson fermions have not been
reported here, they did provide independent confirmation of which signals
were actually contamination that arose due to the twisted mass term.

In all but the $A_2$ channels, the smeared operators produced better fits
than unsmeared operators.
For $A_2^{PC}$, it was found that just
fitting the smeared channel tended to produce failures in the
bootstrap fits which are used to generate the errors of the
parameters.  The reason for
this failure is poor statistics due to a combination of the relatively
few correlators in each $A_2$ channel (no row averaging occurs in a
one-dimensional channel) and the weakness of the expected states in
the channel since $A_2$ couples only to continuum $J\ge 3$ as shown in
Table~\ref{tab:irrepcorrespondence}.  As our motivation for fitting
smeared states is to reduce the influence of high energy states and
since the $A_2$ channels have very few low lying states even in the
unsmeared case, we refit all $A_2$ correlators, both smeared and not,
simultaneously.  This gives sufficient statistics to make our
bootstrap fits converge properly.

Visual inspection of the fits to the $E$ and $T_2$ channels
independently, in each $PC$ combination, revealed that results were
remarkably similar for any state found.  This is to be expected since,
by Table~\ref{tab:irrepcorrespondence}, the lowest angular momentum to
which they both couple is $J=2$.  With the assumption that the few
states found across these channels are of this angular momentum, we do
combined fits of all correlators of $E$ and $T_2$ together for each
$PC$.  In these combined fits the algorithm finds the states to be
statistically the same (i.e.\ it does not bifurcate any of the states),
thus allowing us to conclude that our results are commensurate with
ascribing a value $J=2$ to the common state.\footnote{Technically
  $J=4$ is also a logical possibility since its subduced representation reduces to both irreps
  $E$ and $T_2$ as well.}  Furthermore it allows us to use all the
available data to extract its value.  We thus tabulate results only
for the combined $(E,T_2)$ fits, and not them separately.  Our lattice
spacing is fine enough that given our statistics it is unlikely that
any splitting between $E$ and $T_2$ states for a common $J=2$ state
would be detectable, and it is not.\footnote{Such splitting can be
  observed across continuum subduction channels on coarser lattices.
  See Ref.~\cite{PhysRevD.56.4043}.}  The remaining states in our fits
are assigned their most probable continuum $J$ values based on
Table~\ref{tab:irrepcorrespondence}, namely $J=0$, $1$, and $3$ for
irreps $A_1$, $T_1$, and $A_2$ respectively.\footnote{While $J=3$
  could also appear in $T_1$ and $T_2$ channels as well, the likely
  weakness of such a signal made a combined fitting of these channels
  inappropriate in comparison to the $J=2$ case, given the presence of
  lower $J$ in $T_1$ and $T_2$ channels and the limitations of our
  statistics.}

%%% Local Variables: 
%%% mode: latex
%%% TeX-master: "tmmeson"
%%% End: 

\subsection{\label{subsec:obsoperators}Observations on operators}

To give insight into the nature of the operators which contribute to a
particular state we have produced Table~\ref{tab:irrepcontrib} in
Appendix~\ref{sec:irrepcontrib}.  There one finds the operators which
are found to contribute to each extrapolated state, written in terms
of their local quark and extended gauge field content from which they
are constructed.  The nature of our operators is such that only one
irrep is found in any such product as shown in
Table~\ref{tab:totalreduction} so this labeling uniquely identifies the
operator.

To quantify the significance of an operator's diagonal
correlator to a given channel two additional pieces of information are
given in Table~\ref{tab:irrepcontrib}.  For each such operator the
number of times it was found contributing to the state out of the
twelve possible lattice spacing and quark mass combinations that were
fit is given.  In some cases a small contribution may not be
statistically significant and as such the fitting algorithm may opt
not to give a coefficient for it.  Secondly, the table provides the
largest magnitude of the coefficient $Z$ found for that state, among those
fits that had a coefficient.  These latter values are scaled by
their error so as to give a measure of their significance.  The table
is organized to give the most significant operators contributing to a
channel first in the list.

Several observations may be made from Table~\ref{tab:irrepcontrib}.
For one, operators with a non-scalar gauge structure still contribute
to ground states. For the $\rho$ meson ($T_1^{--}$ ground state) the
greatest contributions come from those operators which have a quark
structure of $T_1^{--}$, either a local operator or with an extended
$A_1^{++}$ gauge field.  However, as can be seen in the table, one
also has a significant overlap with states of the form
$A_1^{-+}\otimes T_1^{+-}$ in which the gauge field is actually
providing the vector nature of the state.\footnote{For the ground
  state pion ($A_1^{-+}$), not shown in Table~\ref{tab:irrepcontrib},
  it is to be remarked that there is similarly a clear ground state
  signal from the $T_1^{--}\otimes T_1^{--}$ operators.} The excited
state $T_1^{*--}$ shows, on the other hand, a much greater relative
contribution from the vector gauge field operators, but still dominant
contributions from the vector quark operators.  For the ground states
$T_1^{++}$ and $T_1^{+-}$ the operators with quark vector structure
dominate, however for their excited $T_1^{*++}$ and $T_1^{*+-}$ states
it is seen that a greater contribution comes from the operators with
vector gauge field structure.
This observation is an example of how lattice studies of the spectrum can
inform discussion of the distinction between hybrid vector mesons and
conventional vector mesons which is made in models
like the flux-tube model~\cite{Close:1995hw}.

Table~\ref{tab:irrepcontrib} also grants further insight into the
geometry of related states.  In the constituent quark model, for
instance, the $a_0$, $a_1$, and $b_1$ states belong to a common P-wave
and hence are considered to share common angular
momentum~\cite{Lacock:1996vy}. If one looks at the main contribution to
these states in their corresponding octahedral channels, $A_1^{++}$,
$T_1^{++}$, and $T_1^{-+}$ respectively, it is observed that after the
local operators the main contributor in each of these channels is of
the form $X^{PC}\otimes T_1^{+-}$, where $X^{PC}$ is whatever local
spin structure is required to produce the correct channel.  This suggests
that just as common spatial angular momenta ($L=1$) are traditionally
considered to unite these channels, so too does similar gauge field
structure.

For those states we identified as $J=2$ with our common fits to $E$
and $T_2$ channels, Table~\ref{tab:irrepcontrib} shows further
affirmation of this identification.  In the continuum, five $J=2$
operators can be formed with the coupling of a $J=1$ quark and $J=1$
gauge field structure.  On the lattice these five states still appear
in a coupling of $T_1$ quark and $T_1$ gauge field, but now in the
subduction of $J=2$ as a two-dimensional $E$ irrep and a
three-dimensional $T_2$ irrep.  As such one expects these $E$ and
$T_2$ operators produced via $T_1\otimes T_1$ corresponding to an
actual $J=2$ state to have similar properties.\footnote{However, since
  one is row averaging over three operators in the $T_2$ case compared
  to two in the $E$ case the correspondence is not expected to be
  identical due to differing statistics.}  A consideration of our
eight $(E,T_2)$ common fits shows that the contributions of such $E$
operators (denoted by $T_1\otimes T_1$) have comparable counts and
significance to their related $T_2$ operators (denoted by $T_1
\overline{\otimes} T_1$) in all cases.  Only in the comparably weak
$2^{++}$ ground state fit, and there only in the neutral channel, is
the pattern less obvious.  For all these states the importance of the
gluonic degrees of freedom is readily apparent, in contradistinction
to the limitations imposed by a simple quark model for two local quarks.

%%% Local Variables:
%%% mode: latex
%%% TeX-master: "tmmeson"
%%% End:

\section{\label{sec:conclusion}Conclusion}

Simulations with twisted mass lattice QCD (tmLQCD) have been used to explore the
spectrum of mesons having all possible values of angular momentum, spin, and
parity.  Numerical results are consistent with those obtained by other authors
using other lattice actions.  As seen in Figure~\ref{fig:final}, we are
still far from seeing the number of mesons claimed experimentally, and
uncertainties are still large, but a viable methodology has been put into place.
In particular, questions related to the appropriateness of tmLQCD for this
physics can now be answered, the usefulness of operators containing local
quark and anti-quark fields can be evaluated, and the valuable qualities of an
evolutionary fitting method can be confirmed.

Because tmLQCD breaks the parity symmetry of the strong interactions,
it is not clear {\em a priori}
whether it can be used to determine the spectrum of hadrons having definite
quantum numbers $J^{PC}$, which appear as $\Lambda^{PC}$ on a discrete lattice.
In the case of both isovector and isoscalar mesons,
we have identified orthogonality relations among certain operators which are
valid in tmLQCD for any chosen twist angle.  We have also emphasized that
some states cannot be separated, such as $\Lambda^{++}, \Lambda^{--}$
in the isovector case and $\Lambda^{++}, \Lambda^{-+}$ in the isoscalar case.
Nevertheless, simultaneous fits to multiple correlators allow the complete
spectrum to be obtained (as proposed in Ref.~\cite{Frezzotti:2003ni}),
and comparison of the isovector spectrum with its
isoscalar counterpart serves to clarify the $PC$ quantum numbers of the states.
During this work, simulations with the Wilson action were performed also in 
order to confirm these claims.

Meson operators that require only one quark propagator, i.e.\ operators with a
common local source for mass-degenerate quark and anti-quark, minimize
computational expense.
Connecting the quark and anti-quark to spatially extended gauge fields
permitted all possible meson quantum numbers, but how well do these
operators overlap with the physical states of QCD?
Our numerical explorations produced signals for both conventional and exotic
mesons.  The strength of each operator's overlap was tabulated, so that
future studies can make informed choices of operators for the various meson
channels.  Gauge field smearing was included at source and sink, while quark
field smearing was performed only at the sink.

The use of an evolutionary fitting algorithm avoided the danger of human bias
during data analysis.  The algorithm was able to fit multiple correlators
simultaneously, with some states shared (or not) across data sets.
The algorithm itself determined how many states were present based on
statistical signficance, and all time steps beyond the source were included in
all fits.  Survival of the fittest was defined to mean minimizing
$\chi^2/n_{dof}$.  The algorithm even served to alert us to a case of
insufficient statistics in one ensemble of lattices; this information appeared
as larger $\chi^2/n_{dof}$ values output from the fit, which could only be
reduced toward unity by increased statistics.

Based on this work, 
we conclude that a detailed study of the full meson spectrum using 
tmLQCD is feasible.  However, to obtain precise results, substantially more
configurations would be required than we have used, and additional classes of
operators should be considered.  Quark smearing at the source, to match the
sink, should improve the signal-to-noise ratio.  With more precise
meson mass determinations at a few $\beta$ values, one could see directly
whether the continuum extrapolation of tmLQCD offers a significant advantage
over other lattice actions, commensurate with the drawback of tmLQCD's lack of
parity conservation.  It would also be interesting to consider the viability of 
using tmLQCD for a thorough study of the baryon spectrum.

%%% Local Variables: 
%%% mode: latex
%%% TeX-master: "tmmeson"
%%% End: 

\appendix

%\input{spectroscopyplotintro}

%randy \clearpage
%randy \twocol{}{\begingroup\singlespacing}

\vspace*{4mm}
\section{\label{sec:irrepcontrib}Operator contribution}

The operators contributing to each extrapolated state are listed in
Table~\ref{tab:irrepcontrib}.  Operators are identified by the direct
    product of quark local irrep and extended gauge field irrep from
    which they are projected,
    \mbox{$\irrepdirac^{\irrepcopy_\diraclower P_\diraclower
        C_\diraclower}\otimes \irreplink^{P_\linklower
        C_\linklower}$}.  Operators without an extended gauge
    structure given are local quark operators.  Quark irreps from the
    second multiplicity have a superscript $2$.  In the case of the
    $(E,T_2)$ fits we designate $T_2$ operators with an
    $\overline{\otimes}$ and $E$ operators with $\otimes$.  In the
    $A_2$ fits where both smeared and unsmeared correlators were fit
    the unsmeared correlators are distinguished similarly with
    $\overline{\otimes}$.  After each operator the number of lattice
    spacing-quark mass channels in which a coefficient was identified
    for the state is given (\mbox{maximum $12$}) followed by the
    maximum value of the coefficient $Z$ divided by its error over
    those channels.

%R Put \setlength{\LTcapwidth}{\textwidth} in the preamble to make
%R the captionwidth span the table in any longtable. for the preprint
%R and {.85\textwidth} in twocolumn mode.
\begin{longtable*}{lcl}
 \caption{\label{tab:irrepcontrib}Table of operators contributing to
   each extrapolated state.}
   \\
   \hline\hline
   \multicolumn{2}{c}{Channel} & \\
   \multicolumn{1}{c}{Irrep(s)}&\multicolumn{1}{c}{C/N} &
   \multicolumn{1}{c}{Contributing Operators} \\ \hline
 \endfirsthead
 \multicolumn{3}{c}%
 {\tablename\ \thetable{} -- Continued from previous page} \\
 \hline\hline
 \multicolumn{2}{c}{Channel} & \\
 \multicolumn{1}{c}{Irrep(s)}&\multicolumn{1}{c}{C/N} & \multicolumn{1}{c}{Contributing Operators} \\ \hline
 \endhead
 \hline\hline
 \multicolumn{3}{r}{Continued on next page} \\
 \endfoot
 \hline\hline
 \endlastfoot
 $A_1^{++}$ & C & $A_1^{++}(12,95.2)$,\ $A_1^{++}\otimes A_1^{++}(12,94.6)$,\ $T_1^{+-}\otimes T_1^{+-}(7,3.97)$,\ $T_1^{2--}\otimes T_1^{--}(2,2.49)$ \\
           & N & $A_1^{++}\otimes A_1^{++}(12,48.4)$,\ $A_1^{++}(12,48.1)$,\ $T_1^{+-}\otimes T_1^{+-}(9,4.34)$,\ $T_1^{2--}\otimes T_1^{--}(3,3.92)$ \\
 $A_1^{+-}$ & C & $A_1^{+-}(12,9.25)$,\ $A_1^{+-}\otimes A_1^{++}(12,9.25)$,\ $T_1^{++}\otimes T_1^{+-}(10,5.74)$,\ $T_1^{2--}\otimes T_1^{-+}(7,4.60)$,\ \\* 
  & & $T_1^{--}\otimes T_1^{-+}(4,4.41)$ \\
           & N & $A_1^{+-}(12,5.26)$,\ $A_1^{+-}\otimes A_1^{++}(12,5.20)$,\ $T_1^{++}\otimes T_1^{+-}(11,4.19)$,\ $T_1^{--}\otimes T_1^{-+}(5,3.11)$,\ \\* 
  & & $T_1^{2--}\otimes T_1^{-+}(2,1.30)$ \\
 $A_1^{*-+}$ & C & $A_1^{-+}(8,29.5)$,\ $A_1^{-+}\otimes A_1^{++}(8,28.9)$,\ $T_1^{--}\otimes T_1^{+-}(8,11.3)$,\ $T_1^{2--}\otimes T_1^{+-}(8,3.57)$,\ \\* 
  & & $T_1^{+-}\otimes T_1^{--}(4,9.21)$,\ $A_1^{2-+}(4,3.11)$,\ $T_1^{++}\otimes T_1^{-+}(4,0.92)$,\ $A_1^{2-+}\otimes A_1^{++}(3,3.46)$ \\
           & N & $A_1^{-+}\otimes A_1^{++}(12,26.9)$,\ $A_1^{-+}(12,26.9)$,\ $A_1^{2-+}\otimes A_1^{++}(10,5.53)$,\ $T_1^{2--}\otimes T_1^{+-}(8,3.26)$,\ \\* 
  & & $A_1^{2-+}(7,5.58)$,\ $T_1^{+-}\otimes T_1^{--}(6,4.38)$,\ $T_1^{--}\otimes T_1^{+-}(5,5.08)$,\ $T_1^{++}\otimes T_1^{-+}(4,4.92)$ \\
 $T_1^{++}$ & C & $T_1^{++}\otimes A_1^{++}(11,36.8)$,\ $T_1^{++}(11,36.7)$,\ $T_1^{+-}\otimes T_1^{+-}(7,2.47)$,\ $T_1^{+-}\otimes E^{+-}(7,2.00)$,\ \\* 
  & & $T_1^{--}\otimes T_2^{--}(3,2.14)$,\ $T_1^{2--}\otimes T_2^{--}(3,2.12)$,\ $A_1^{+-}\otimes T_1^{+-}(3,1.32)$,\ $T_1^{2--}\otimes T_1^{--}(2,1.45)$,\ \\* 
  & & $T_1^{--}\otimes T_1^{--}(2,1.39)$,\ $T_1^{++}\otimes T_2^{++}(2,0.60)$,\ $T_1^{++}\otimes E^{++}(1,1.42)$ \\
           & N & $T_1^{++}\otimes A_1^{++}(12,61.8)$,\ $T_1^{++}(12,50.0)$,\ $T_1^{+-}\otimes T_1^{+-}(8,5.43)$,\ $T_1^{+-}\otimes E^{+-}(7,3.54)$,\ \\* 
  & & $T_1^{2--}\otimes T_1^{--}(7,1.69)$,\ $A_1^{+-}\otimes T_1^{+-}(6,1.97)$,\ $T_1^{2--}\otimes T_2^{--}(5,4.16)$,\ $T_1^{--}\otimes T_1^{--}(5,2.98)$,\ \\* 
  & & $T_1^{--}\otimes T_2^{--}(5,0.80)$,\ $T_1^{++}\otimes E^{++}(2,0.69)$,\ $T_1^{++}\otimes T_2^{++}(1,0.77)$,\ $A_1^{-+}\otimes T_1^{-+}(1,0.03)$ \\
 $T_1^{*++}$ & C & $A_1^{+-}\otimes T_1^{+-}(12,6.56)$,\ $T_1^{+-}\otimes T_1^{+-}(11,6.70)$,\ $T_1^{++}\otimes A_1^{++}(8,9.12)$,\ $T_1^{++}(7,9.46)$,\ \\* 
  & & $T_1^{--}\otimes T_1^{--}(6,5.24)$,\ $A_1^{-+}\otimes T_1^{-+}(6,2.76)$,\ $T_1^{2--}\otimes T_2^{--}(6,2.55)$,\ $T_1^{--}\otimes T_2^{--}(6,1.96)$,\ \\* 
  & & $A_1^{2-+}\otimes T_1^{-+}(5,4.11)$,\ $T_1^{++}\otimes T_2^{++}(5,2.46)$,\ $T_1^{++}\otimes E^{++}(5,1.68)$,\ $T_1^{2--}\otimes T_1^{--}(3,4.45)$,\ \\* 
  & & $T_1^{+-}\otimes E^{+-}(2,1.89)$ \\
           & N & $A_1^{+-}\otimes T_1^{+-}(12,6.02)$,\ $T_1^{+-}\otimes T_1^{+-}(11,6.80)$,\ $T_1^{++}\otimes A_1^{++}(9,3.79)$,\ $T_1^{++}(9,3.53)$,\ \\* 
  & & $A_1^{2-+}\otimes T_1^{-+}(9,2.61)$,\ $T_1^{--}\otimes T_2^{--}(8,2.89)$,\ $A_1^{-+}\otimes T_1^{-+}(8,2.24)$,\ $T_1^{2--}\otimes T_2^{--}(7,1.31)$,\ \\* 
  & & $T_1^{2--}\otimes T_1^{--}(5,2.71)$,\ $T_1^{++}\otimes E^{++}(5,2.67)$,\ $T_1^{++}\otimes T_2^{++}(5,1.86)$,\ $T_1^{--}\otimes T_1^{--}(4,2.47)$,\ \\* 
  & & $T_1^{+-}\otimes E^{+-}(4,1.05)$ \\
 $T_1^{+-}$ & C & $T_1^{+-}\otimes A_1^{++}(12,38.6)$,\ $T_1^{+-}(12,35.3)$,\ $T_1^{++}\otimes T_1^{+-}(8,6.19)$,\ $A_1^{++}\otimes T_1^{+-}(7,2.78)$,\ \\* 
  & & $A_1^{-+}\otimes T_1^{--}(5,2.56)$,\ $T_1^{++}\otimes E^{+-}(5,2.41)$,\ $T_1^{--}\otimes T_2^{-+}(4,3.83)$,\ $T_1^{2--}\otimes T_1^{-+}(4,1.59)$,\ \\* 
  & & $A_1^{2-+}\otimes T_1^{--}(4,1.41)$,\ $T_1^{2--}\otimes T_2^{-+}(2,1.64)$,\ $T_1^{+-}\otimes T_2^{++}(2,0.98)$ \\
           & N & $T_1^{+-}\otimes A_1^{++}(12,29.9)$,\ $T_1^{+-}(12,29.6)$,\ $T_1^{++}\otimes T_1^{+-}(8,4.67)$,\ $A_1^{++}\otimes T_1^{+-}(8,3.92)$,\ \\* 
  & & $T_1^{2--}\otimes T_1^{-+}(3,3.44)$,\ $A_1^{2-+}\otimes T_1^{--}(3,2.34)$,\ $T_1^{++}\otimes E^{+-}(3,2.30)$,\ $T_1^{--}\otimes T_1^{-+}(3,1.16)$,\ \\* 
  & & $T_1^{+-}\otimes T_2^{++}(2,1.16)$,\ $A_1^{-+}\otimes T_1^{--}(2,0.97)$,\ $T_1^{--}\otimes T_2^{-+}(1,0.96)$,\ $T_1^{2--}\otimes T_2^{-+}(1,0.67)$ \\
%\twocol{}{\newpage}
 $T_1^{*+-}$ & C & $T_1^{++}\otimes T_1^{+-}(10,7.08)$,\ $A_1^{++}\otimes T_1^{+-}(9,5.89)$,\ $T_1^{+-}\otimes A_1^{++}(9,4.79)$,\ $T_1^{+-}(9,3.68)$,\ \\* 
  & & $T_1^{--}\otimes T_1^{-+}(7,2.83)$,\ $A_1^{2-+}\otimes T_1^{--}(6,4.69)$,\ $A_1^{-+}\otimes T_1^{--}(5,4.93)$,\ $T_1^{++}\otimes E^{+-}(4,2.22)$,\ \\* 
  & & $T_1^{2--}\otimes T_2^{-+}(3,3.24)$,\ $T_1^{--}\otimes T_2^{-+}(3,2.48)$,\ $T_1^{2--}\otimes T_1^{-+}(3,2.17)$,\ $T_1^{+-}\otimes E^{++}(2,2.88)$,\ \\* 
  & & $T_1^{+-}\otimes T_2^{++}(2,0.88)$ \\
           & N & $T_1^{++}\otimes T_1^{+-}(10,7.28)$,\ $A_1^{++}\otimes T_1^{+-}(9,7.23)$,\ $T_1^{+-}(6,9.53)$,\ $T_1^{+-}\otimes A_1^{++}(6,9.52)$,\ \\* 
  & & $A_1^{2-+}\otimes T_1^{--}(5,5.36)$,\ $T_1^{--}\otimes T_1^{-+}(5,5.30)$,\ $A_1^{-+}\otimes T_1^{--}(4,4.53)$,\ $T_1^{2--}\otimes T_2^{-+}(4,4.02)$,\ \\* 
  & & $T_1^{--}\otimes T_2^{-+}(4,3.27)$,\ $T_1^{2--}\otimes T_1^{-+}(4,2.82)$,\ $T_1^{++}\otimes E^{+-}(4,2.59)$,\ $T_1^{+-}\otimes E^{++}(3,3.36)$,\ \\* 
  & & $T_1^{+-}\otimes T_2^{++}(3,1.73)$ \\
 $T_1^{-+}$ & C & $T_1^{2--}\otimes T_1^{+-}(12,7.54)$,\ $T_1^{--}\otimes T_1^{+-}(12,7.31)$,\ $T_1^{+-}\otimes T_2^{--}(4,3.99)$,\ $T_1^{2--}\otimes E^{+-}(4,2.73)$,\ \\* 
  & & $T_1^{--}\otimes E^{+-}(4,2.56)$,\ $A_1^{+-}\otimes T_1^{--}(4,1.66)$,\ $T_1^{++}\otimes T_2^{-+}(3,3.67)$,\ $A_1^{++}\otimes T_1^{-+}(3,1.81)$,\ \\* 
  & & $T_1^{+-}\otimes T_1^{--}(3,1.36)$,\ $T_1^{++}\otimes T_1^{-+}(2,2.70)$ \\
           & N & $T_1^{--}\otimes T_1^{+-}(11,7.29)$,\ $T_1^{2--}\otimes T_1^{+-}(11,7.09)$,\ $T_1^{+-}\otimes T_1^{--}(9,3.45)$,\ $T_1^{++}\otimes T_1^{-+}(7,3.03)$,\ \\* 
  & & $A_1^{++}\otimes T_1^{-+}(6,2.78)$,\ $A_1^{+-}\otimes T_1^{--}(4,4.50)$,\ $T_1^{++}\otimes T_2^{-+}(4,3.87)$,\ $T_1^{2--}\otimes E^{+-}(4,2.34)$,\ \\* 
  & & $T_1^{--}\otimes E^{+-}(4,1.97)$,\ $T_1^{+-}\otimes T_2^{--}(3,3.02)$ \\
%\twocol{\newpage}{}
 $T_1^{--}$ & C & $T_1^{2--}\otimes A_1^{++}(12,64.1)$,\ $T_1^{2--}(12,64.1)$,\ $T_1^{--}(12,43.7)$,\ $T_1^{--}\otimes A_1^{++}(12,43.7)$,\ \\* 
  & & $A_1^{2-+}\otimes T_1^{+-}(7,7.34)$,\ $A_1^{-+}\otimes T_1^{+-}(6,4.84)$,\ $T_1^{2--}\otimes E^{++}(4,2.43)$,\ $T_1^{--}\otimes T_2^{++}(4,0.95)$,\ \\* 
  & & $T_1^{++}\otimes T_1^{--}(3,1.94)$,\ $T_1^{++}\otimes T_2^{--}(3,1.85)$,\ $T_1^{2--}\otimes T_2^{++}(3,1.50)$,\ $A_1^{++}\otimes T_1^{--}(3,1.29)$,\ \\* 
  & & $T_1^{--}\otimes E^{++}(3,0.84)$,\ $T_1^{+-}\otimes T_1^{-+}(2,2.74)$,\ $A_1^{+-}\otimes T_1^{-+}(1,1.32)$ \\
           & N & $T_1^{--}(12,107.)$,\ $T_1^{--}\otimes A_1^{++}(12,107.)$,\ $T_1^{2--}(10,25.7)$,\ $T_1^{2--}\otimes A_1^{++}(10,25.1)$,\ \\* 
  & & $A_1^{-+}\otimes T_1^{+-}(9,8.06)$,\ $A_1^{2-+}\otimes T_1^{+-}(8,7.07)$,\ $T_1^{++}\otimes T_2^{--}(8,1.35)$,\ $A_1^{++}\otimes T_1^{--}(6,3.08)$,\ \\* 
  & & $T_1^{+-}\otimes T_1^{-+}(5,2.42)$,\ $T_1^{++}\otimes T_1^{--}(5,2.41)$,\ $T_1^{--}\otimes E^{++}(5,1.22)$,\ $T_1^{2--}\otimes E^{++}(3,1.87)$,\ \\* 
  & & $A_1^{+-}\otimes T_1^{-+}(3,0.41)$,\ $T_1^{--}\otimes T_2^{++}(2,1.86)$,\ $T_1^{+-}\otimes T_2^{-+}(2,1.83)$,\ $T_1^{2--}\otimes T_2^{++}(2,1.00)$ \\
 $T_1^{*--}$ & C & $T_1^{--}(12,22.0)$,\ $T_1^{--}\otimes A_1^{++}(12,21.9)$,\ $T_1^{2--}(11,17.4)$,\ $T_1^{2--}\otimes A_1^{++}(11,17.3)$,\ \\* 
  & & $A_1^{-+}\otimes T_1^{+-}(11,4.39)$,\ $A_1^{2-+}\otimes T_1^{+-}(10,7.17)$,\ $T_1^{++}\otimes T_1^{--}(7,2.83)$,\ $A_1^{++}\otimes T_1^{--}(7,2.81)$,\ \\* 
  & & $T_1^{--}\otimes E^{++}(7,1.63)$,\ $T_1^{++}\otimes T_2^{--}(5,2.03)$,\ $T_1^{--}\otimes T_2^{++}(4,1.05)$,\ $T_1^{+-}\otimes T_1^{-+}(2,1.57)$,\ \\* 
  & & $T_1^{+-}\otimes T_2^{-+}(2,1.43)$,\ $A_1^{+-}\otimes T_1^{-+}(2,0.60)$,\ $T_1^{2--}\otimes E^{++}(1,1.62)$,\ $T_1^{2--}\otimes T_2^{++}(1,1.01)$ \\
           & N & $T_1^{2--}\otimes A_1^{++}(12,46.9)$,\ $T_1^{2--}(12,45.3)$,\ $A_1^{2-+}\otimes T_1^{+-}(10,4.82)$,\ $T_1^{--}\otimes A_1^{++}(9,19.3)$,\ \\* 
  & & $T_1^{--}(9,18.9)$,\ $A_1^{++}\otimes T_1^{--}(7,3.16)$,\ $A_1^{-+}\otimes T_1^{+-}(6,4.15)$,\ $T_1^{++}\otimes T_1^{--}(6,4.07)$,\ \\* 
  & & $T_1^{2--}\otimes E^{++}(3,1.48)$,\ $T_1^{++}\otimes T_2^{--}(3,1.30)$,\ $T_1^{+-}\otimes T_2^{-+}(3,1.25)$,\ $A_1^{+-}\otimes T_1^{-+}(3,1.02)$,\ \\* 
  & & $T_1^{2--}\otimes T_2^{++}(1,1.72)$ \\
 $(E,T_2)^{++}$ & C & $T_1^{++}\overline{\otimes} E^{++}(7,1.09)$,\ $T_1^{2--}\otimes T_2^{--}(6,1.80)$,\ $T_1^{2--}\otimes T_1^{--}(6,1.24)$,\ $T_1^{--}\overline{\otimes} T_2^{--}(6,1.13)$,\ \\* 
  & & $T_1^{2--}\overline{\otimes} T_1^{--}(4,1.79)$,\ $T_1^{+-}\overline{\otimes} T_1^{+-}(4,1.24)$,\ $T_1^{--}\otimes T_2^{--}(4,1.07)$,\ $A_1^{+-}\otimes E^{+-}(4,0.99)$,\ \\* 
  & & $T_1^{+-}\otimes T_1^{+-}(3,1.39)$,\ $A_1^{-+}\overline{\otimes} T_2^{-+}(3,0.99)$,\ $T_1^{+-}\overline{\otimes} A_2^{+-}(3,0.88)$,\ $T_1^{--}\otimes T_1^{--}(2,0.74)$,\ \\* 
  & & $A_1^{++}\overline{\otimes} T_2^{++}(1,1.16)$,\ $T_1^{--}\overline{\otimes} T_1^{--}(1,0.55)$,\ $T_1^{2--}\overline{\otimes} T_2^{--}(1,0.38)$,\ $T_1^{+-}\overline{\otimes} E^{+-}(1,0.04)$ \\
           & N & $T_1^{+-}\overline{\otimes} E^{+-}(6,1.35)$,\ $T_1^{--}\overline{\otimes} T_2^{--}(5,1.89)$,\ $T_1^{+-}\overline{\otimes} T_1^{+-}(4,1.55)$,\ $T_1^{--}\otimes T_1^{--}(4,1.48)$,\ \\* 
  & & $T_1^{2--}\overline{\otimes} T_2^{--}(4,1.42)$,\ $T_1^{2--}\overline{\otimes} T_1^{--}(3,2.20)$,\ $T_1^{+-}\otimes T_1^{+-}(3,1.28)$,\ $T_1^{++}\overline{\otimes} E^{++}(3,1.04)$,\ \\* 
  & & $T_1^{+-}\overline{\otimes} A_2^{+-}(3,0.89)$,\ $A_1^{+-}\otimes E^{+-}(2,1.31)$,\ $T_1^{--}\otimes T_2^{--}(2,0.94)$,\ $T_1^{2--}\otimes T_2^{--}(2,0.93)$,\ \\* 
  & & $A_1^{++}\overline{\otimes} T_2^{++}(2,0.69)$,\ $T_1^{++}\otimes T_2^{++}(1,0.74)$ \\
%\twocol{}{\newpage}
 $(E,T_2)^{*++}$ & C & $T_1^{+-}\overline{\otimes} T_1^{+-}(11,5.66)$,\ $T_1^{+-}\otimes T_1^{+-}(11,4.66)$,\ $A_1^{++}\otimes E^{++}(7,2.06)$,\ $A_1^{2-+}\overline{\otimes} T_2^{-+}(6,2.27)$,\ \\* 
  & & $T_1^{--}\otimes T_2^{--}(6,2.06)$,\ $T_1^{2--}\otimes T_2^{--}(6,2.02)$,\ $T_1^{--}\overline{\otimes} T_2^{--}(6,1.75)$,\ $A_1^{++}\overline{\otimes} T_2^{++}(6,1.52)$,\ \\* 
  & & $T_1^{++}\overline{\otimes} E^{++}(4,2.27)$,\ $T_1^{2--}\otimes T_1^{--}(4,2.23)$,\ $T_1^{2--}\overline{\otimes} T_1^{--}(4,2.20)$,\ $A_1^{-+}\overline{\otimes} T_2^{-+}(4,2.20)$,\ \\* 
  & & $T_1^{--}\overline{\otimes} T_1^{--}(4,2.19)$,\ $T_1^{--}\otimes T_1^{--}(4,2.05)$,\ $T_1^{2--}\overline{\otimes} T_2^{--}(4,1.96)$,\ $T_1^{+-}\overline{\otimes} E^{+-}(4,1.82)$,\ \\* 
  & & $T_1^{++}\overline{\otimes} T_2^{++}(4,1.71)$,\ $T_1^{+-}\overline{\otimes} A_2^{+-}(4,1.61)$,\ $T_1^{++}\otimes T_2^{++}(4,1.51)$,\ $A_1^{+-}\otimes E^{+-}(4,1.12)$ \\
           & N & $T_1^{+-}\overline{\otimes} T_1^{+-}(12,5.50)$,\ $T_1^{+-}\otimes T_1^{+-}(12,4.73)$,\ $T_1^{--}\overline{\otimes} T_2^{--}(8,3.08)$,\ $T_1^{2--}\otimes T_2^{--}(7,3.85)$,\ \\* 
  & & $T_1^{--}\otimes T_2^{--}(7,3.50)$,\ $T_1^{2--}\overline{\otimes} T_2^{--}(7,2.96)$,\ $A_1^{++}\overline{\otimes} T_2^{++}(7,1.92)$,\ $T_1^{--}\overline{\otimes} T_1^{--}(5,3.85)$,\ \\* 
  & & $A_1^{+-}\otimes E^{+-}(5,2.91)$,\ $T_1^{2--}\otimes T_1^{--}(4,4.34)$,\ $T_1^{2--}\overline{\otimes} T_1^{--}(4,4.23)$,\ $A_1^{-+}\overline{\otimes} T_2^{-+}(4,3.69)$,\ \\* 
  & & $A_1^{2-+}\overline{\otimes} T_2^{-+}(4,3.54)$,\ $T_1^{+-}\overline{\otimes} E^{+-}(4,3.03)$,\ $T_1^{+-}\overline{\otimes} A_2^{+-}(4,1.59)$,\ $A_1^{++}\otimes E^{++}(3,3.59)$,\ \\* 
  & & $T_1^{--}\otimes T_1^{--}(3,2.30)$,\ $T_1^{++}\otimes T_2^{++}(3,1.92)$,\ $T_1^{++}\overline{\otimes} T_2^{++}(2,1.34)$,\ $T_1^{++}\overline{\otimes} E^{++}(1,2.04)$ \\
 $(E,T_2)^{+-}$ & C & $T_1^{++}\overline{\otimes} T_1^{+-}(11,4.69)$,\ $T_1^{++}\otimes T_1^{+-}(11,4.59)$,\ $T_1^{2--}\otimes T_1^{-+}(7,3.96)$,\ $T_1^{--}\overline{\otimes} T_1^{-+}(6,3.64)$,\ \\* 
  & & $T_1^{--}\otimes T_1^{-+}(5,3.81)$,\ $A_1^{2-+}\overline{\otimes} T_2^{--}(5,3.50)$,\ $A_1^{-+}\overline{\otimes} T_2^{--}(5,3.34)$,\ $T_1^{2--}\overline{\otimes} T_1^{-+}(4,3.42)$,\ \\* 
  & & $T_1^{--}\otimes T_2^{-+}(4,3.35)$,\ $T_1^{--}\overline{\otimes} T_2^{-+}(4,3.06)$,\ $T_1^{+-}\otimes T_2^{++}(4,2.78)$,\ $T_1^{++}\overline{\otimes} A_2^{+-}(4,1.59)$,\ \\* 
  & & $T_1^{+-}\overline{\otimes} E^{++}(3,3.98)$,\ $A_1^{+-}\otimes E^{++}(3,3.91)$,\ $T_1^{2--}\otimes T_2^{-+}(3,3.52)$,\ $T_1^{2--}\overline{\otimes} T_2^{-+}(3,2.95)$,\ \\* 
  & & $A_1^{+-}\overline{\otimes} T_2^{++}(3,2.56)$,\ $A_1^{++}\otimes E^{+-}(3,2.31)$,\ $T_1^{+-}\overline{\otimes} T_2^{++}(3,2.09)$,\ $T_1^{++}\overline{\otimes} E^{+-}(3,1.49)$ \\
           & N & $T_1^{++}\otimes T_1^{+-}(11,4.83)$,\ $T_1^{++}\overline{\otimes} T_1^{+-}(11,4.63)$,\ $T_1^{2--}\otimes T_1^{-+}(8,3.82)$,\ $T_1^{--}\otimes T_1^{-+}(7,3.56)$,\ \\* 
  & & $T_1^{--}\overline{\otimes} T_1^{-+}(6,3.58)$,\ $A_1^{2-+}\overline{\otimes} T_2^{--}(4,3.85)$,\ $A_1^{-+}\overline{\otimes} T_2^{--}(4,3.31)$,\ $T_1^{+-}\overline{\otimes} T_2^{++}(4,2.81)$,\ \\* 
  & & $A_1^{+-}\otimes E^{++}(3,4.00)$,\ $T_1^{2--}\otimes T_2^{-+}(3,3.93)$,\ $T_1^{+-}\overline{\otimes} E^{++}(3,3.89)$,\ $T_1^{2--}\overline{\otimes} T_1^{-+}(3,3.86)$,\ \\* 
  & & $T_1^{--}\otimes T_2^{-+}(3,3.75)$,\ $T_1^{--}\overline{\otimes} T_2^{-+}(3,3.37)$,\ $T_1^{2--}\overline{\otimes} T_2^{-+}(3,3.22)$,\ $A_1^{+-}\overline{\otimes} T_2^{++}(3,3.15)$,\ \\* 
  & & $T_1^{+-}\otimes T_2^{++}(3,2.81)$,\ $T_1^{++}\overline{\otimes} A_2^{+-}(3,2.13)$,\ $A_1^{++}\otimes E^{+-}(3,2.06)$,\ $T_1^{++}\overline{\otimes} E^{+-}(3,1.56)$ \\
 $(E,T_2)^{-+}$ & C & $T_1^{2--}\overline{\otimes} T_1^{+-}(12,6.03)$,\ $T_1^{--}\otimes T_1^{+-}(12,5.94)$,\ $T_1^{--}\overline{\otimes} T_1^{+-}(12,5.72)$,\ $T_1^{2--}\otimes T_1^{+-}(12,5.43)$,\ \\* 
  & & $T_1^{++}\overline{\otimes} T_1^{-+}(10,3.32)$,\ $T_1^{+-}\otimes T_2^{--}(8,4.05)$,\ $T_1^{+-}\overline{\otimes} T_2^{--}(7,3.99)$,\ $T_1^{++}\overline{\otimes} T_2^{-+}(6,3.61)$,\ \\* 
  & & $A_1^{+-}\overline{\otimes} T_2^{--}(5,4.17)$,\ $T_1^{++}\otimes T_1^{-+}(5,3.56)$,\ $T_1^{+-}\otimes T_1^{--}(4,5.15)$,\ $A_1^{-+}\otimes E^{++}(4,4.50)$,\ \\* 
  & & $T_1^{+-}\overline{\otimes} T_1^{--}(4,4.33)$,\ $A_1^{++}\overline{\otimes} T_2^{-+}(4,3.93)$,\ $A_1^{2-+}\otimes E^{++}(4,3.76)$,\ $T_1^{++}\otimes T_2^{-+}(4,3.65)$,\ \\* 
  & & $A_1^{2-+}\overline{\otimes} T_2^{++}(4,3.04)$,\ $T_1^{2--}\overline{\otimes} E^{+-}(4,2.36)$,\ $T_1^{--}\overline{\otimes} E^{+-}(4,1.88)$,\ $T_1^{--}\overline{\otimes} A_2^{+-}(4,1.76)$,\ \\* 
  & & $A_1^{-+}\overline{\otimes} T_2^{++}(4,1.73)$,\ $T_1^{2--}\overline{\otimes} A_2^{+-}(4,1.64)$ \\
           & N & $T_1^{--}\overline{\otimes} T_1^{+-}(11,5.68)$,\ $T_1^{2--}\overline{\otimes} T_1^{+-}(11,5.63)$,\ $T_1^{--}\otimes T_1^{+-}(11,4.79)$,\ $T_1^{2--}\otimes T_1^{+-}(10,4.74)$,\ \\* 
  & & $T_1^{+-}\otimes T_2^{--}(7,3.61)$,\ $T_1^{++}\overline{\otimes} T_1^{-+}(5,3.88)$,\ $T_1^{++}\otimes T_1^{-+}(5,3.40)$,\ $T_1^{++}\overline{\otimes} T_2^{-+}(5,2.71)$,\ \\* 
  & & $A_1^{+-}\overline{\otimes} T_2^{--}(4,3.91)$,\ $T_1^{+-}\overline{\otimes} T_2^{--}(4,3.33)$,\ $T_1^{++}\otimes T_2^{-+}(4,2.49)$,\ $A_1^{-+}\overline{\otimes} T_2^{++}(4,2.17)$,\ \\* 
  & & $T_1^{--}\overline{\otimes} A_2^{+-}(4,1.92)$,\ $A_1^{-+}\otimes E^{++}(3,4.60)$,\ $T_1^{+-}\otimes T_1^{--}(3,4.42)$,\ $T_1^{2--}\overline{\otimes} E^{+-}(3,2.47)$,\ \\* 
  & & $A_1^{2-+}\otimes E^{++}(2,4.70)$,\ $T_1^{+-}\overline{\otimes} T_1^{--}(2,3.94)$,\ $A_1^{++}\overline{\otimes} T_2^{-+}(2,3.57)$,\ $A_1^{2-+}\overline{\otimes} T_2^{++}(2,3.11)$,\ \\* 
  & & $T_1^{--}\overline{\otimes} E^{+-}(2,2.44)$,\ $T_1^{2--}\overline{\otimes} A_2^{+-}(2,1.63)$ \\
%\twocol{\newpage}{}
 $A_2^{++}$ & C & $T_1^{--}\overline{\otimes} T_2^{--}(7,1.25)$,\ $T_1^{2--}\overline{\otimes} T_2^{--}(7,0.89)$,\ $T_1^{2--}\otimes T_2^{--}(5,1.53)$,\ $T_1^{++}\overline{\otimes} T_2^{++}(2,0.72)$,\ \\* 
  & & $T_1^{++}\otimes T_2^{++}(1,1.27)$,\ $A_1^{+-}\otimes A_2^{+-}(1,0.50)$ \\
           & N & $T_1^{--}\overline{\otimes} T_2^{--}(8,0.95)$,\ $A_1^{+-}\overline{\otimes} A_2^{+-}(6,1.13)$,\ $T_1^{2--}\otimes T_2^{--}(5,1.82)$,\ $T_1^{++}\overline{\otimes} T_2^{++}(5,1.15)$,\ \\* 
  & & $T_1^{2--}\overline{\otimes} T_2^{--}(4,0.00)$,\ $T_1^{++}\otimes T_2^{++}(2,1.77)$,\ $A_1^{+-}\otimes A_2^{+-}(1,0.30)$ \\
 $A_2^{-+}$ & C & $T_1^{+-}\otimes T_2^{--}(6,1.43)$,\ $T_1^{+-}\overline{\otimes} T_2^{--}(6,0.78)$,\ $T_1^{++}\otimes T_2^{-+}(4,2.23)$,\ $T_1^{++}\overline{\otimes} T_2^{-+}(1,1.09)$ \\
           & N & $T_1^{+-}\overline{\otimes} T_2^{--}(8,1.18)$,\ $T_1^{+-}\otimes T_2^{--}(7,2.24)$,\ $T_1^{++}\otimes T_2^{-+}(4,1.94)$,\ $T_1^{++}\overline{\otimes} T_2^{-+}(1,0.97)$ \\
\end{longtable*}
%randy \twocol{}{\endgroup\newpage}
%}

\begin{acknowledgments}
This work was supported in part by the Natural Sciences and
Engineering Research Council (NSERC) of Canada, the Canada Research
Chairs Program, the Canada Foundation for Innovation,
and the Government of Saskatchewan.
\end{acknowledgments}

\bibliography{tmmeson}

\end{document}